\documentclass[acmtog]{acmart} 
\acmSubmissionID{463}

\usepackage{booktabs}

\setcitestyle{nosort,square} 

\usepackage{hyperref}
\usepackage{amsmath}
\usepackage{graphicx}
\usepackage{comment}
\usepackage{kbordermatrix}
\usepackage{multirow,bigdelim}

\usepackage{array}
\usepackage{wrapfig}

\usepackage{csvsimple}
\usepackage{adjustbox}
\usepackage[percent]{overpic}

\newcommand{\etal}{et al.}
\newcommand{\bbb}{{\bf B}}
\newcommand{\bbf}{{\bf F}}
\newcommand{\bbv}{{\bf V}}
\newcommand{\bbo}{{\bf O}}
\newcommand{\bbw}{{\bf W}}
\newcommand{\bbr}{{\bf R}}
\newcommand{\bbt}{{\bf T}}
\newcommand{\Loss}{\mathcal{L}}

\usepackage{dsfont}
\newcommand{\R}{\mathds{R}} 

\usepackage{pgfplots}
\pgfplotsset{compat=newest}
\usepgfplotslibrary{groupplots}
\usepgfplotslibrary{dateplot}

\usepackage[bottom]{footmisc}
\usepackage{color}
\usepackage{soul}
\usepackage{breqn}
\usepackage{booktabs}
\usepackage{tabularx}
\usepackage{rotating}

\definecolor{purple}{rgb}{0.65,0,0.65}
\definecolor{blue}{rgb}{0, 0.2, 0.8}
\definecolor{orange}{rgb}{0.6, 0.6, 0}
\definecolor{red}{rgb}{0.8, 0.2, 0.2}
\definecolor{magenta}{rgb}{0.5, 0.0, 1.0}
\definecolor{black}{rgb}{0.0, 0.0, 0.0}
\definecolor{cyan}{rgb}{0, 0.65, 0.65}
\definecolor{olive}{rgb}{0.2, 0.6, 0.5}
\usepackage{xcolor}

\newif\ifdraft
\drafttrue

\ifdraft
\newcommand{\rhc}[1]{{\color{magenta}\textbf{Rana:} #1}}
\newcommand{\kac}[1]{{\color{orange}\textbf{Kfir:} #1}}
\newcommand{\plc}[1]{{\color{blue}\textbf{Peizhuo:} #1}}
\newcommand{\OSH}[1]{{\color{purple}\textbf{Olga:} #1}}

\else
\newcommand{\rhc}[1]{}
\newcommand{\kac}[1]{}
\newcommand{\plc}[1]{}
\newcommand{\OSH}[1]{}

\fi

\acmJournal{TOG}

\setcopyright{acmcopyright}\acmJournal{TOG}
\acmYear{2021}\acmVolume{40}\acmNumber{4}\acmArticle{1}\acmMonth{8}
\acmDOI{10.1145/3450626.3459852}

\begin{document}

\title{Learning Skeletal Articulations with Neural Blend Shapes}

\author{Peizhuo Li}
\affiliation{\institution{CFCS, Peking University \& AICFVE, Beijing Film Academy}}

\author{Kfir Aberman}
\affiliation{\institution{Google Research}}

\author{Rana Hanocka}
\affiliation{%
    \institution{Tel-Aviv University}
}

\author{Libin Liu}
\affiliation{\institution{CFCS, Peking University }}

\author{Olga Sorkine-Hornung}
\affiliation{\institution{ETH Zurich \& AICFVE, Beijing Film Academy}}

\author{Baoquan Chen}
\affiliation{\institution{CFCS, Peking University \& AICFVE, Beijing Film Academy}}
\authornote{corresponding author}

\authorsaddresses{Authors' addresses: Peizhuo Li, peizhuo@pku.edu.cn; Kfir Aberman, \mbox{kfiraberman@gmail.com}; Rana Hanocka, ranahanocka@gmail.com; Libin Liu, libin.liu@pku.edu.cn; Olga Sorkine-Hornung, sorkine@inf.ethz.ch; Baoquan Chen, \mbox{baoquan@pku.edu.cn}}

\renewcommand\shortauthors{Li, P. et al.}

\begin{abstract}
Animating a newly designed character using motion capture (mocap) data is a long standing problem in computer animation. A key consideration is the skeletal structure that should correspond to the available mocap data, and the shape deformation in the joint regions, which often requires a tailored, pose-specific refinement. In this work, we develop a neural technique for articulating 3D characters using enveloping with a pre-defined skeletal structure which produces high quality pose dependent deformations. Our framework learns to rig and skin characters with the same articulation structure (\textit{e.g.,} bipeds or quadrupeds), and builds the desired skeleton hierarchy into the network architecture. 
\emph{Furthermore}, we propose \emph{neural blend shapes}~--~a set of corrective pose-dependent shapes which improve the deformation quality in the joint regions in order to address the notorious artifacts resulting from standard rigging and skinning.
Our system estimates neural blend shapes for input meshes with arbitrary connectivity, as well as weighting coefficients which are conditioned on the input joint rotations. 
Unlike recent deep learning techniques which supervise the network with ground-truth rigging and skinning parameters, our approach does not assume that the training data has a specific underlying deformation model. Instead, during training, the network observes deformed shapes and learns to infer the corresponding rig, skin and blend shapes using \textit{indirect supervision}.
 During inference, we demonstrate that our network generalizes to unseen characters with arbitrary mesh connectivity, including unrigged characters built by 3D artists. Conforming to standard skeletal animation models enables {direct} plug-and-play in standard animation software, as well as game engines.
\end{abstract}

\begin{teaserfigure}
\centering
\includegraphics[width=\linewidth]{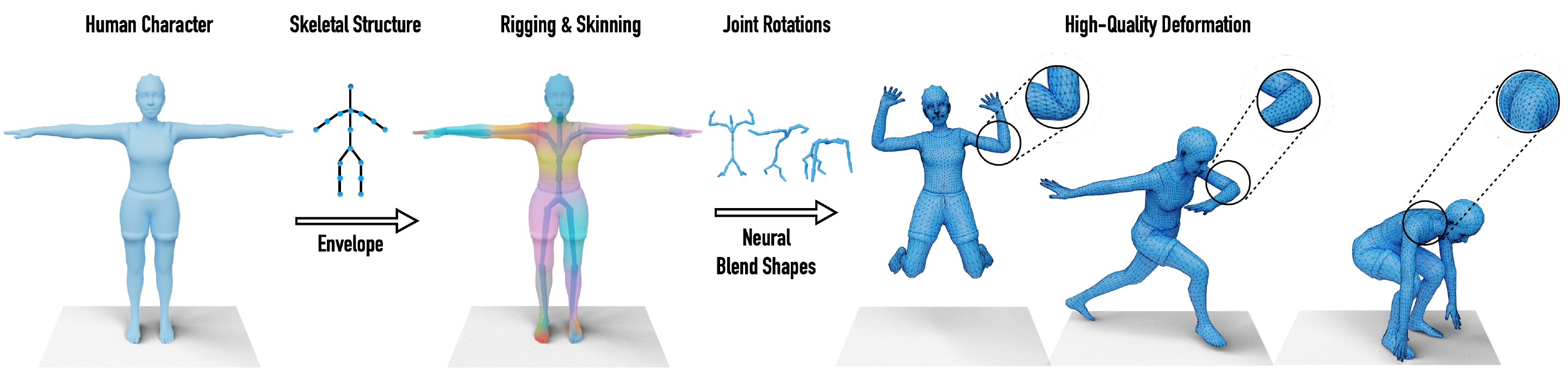} 
\caption{Our neural skeletal articulation network learns to rig and skin an input character with arbitrary connectivity, and generates neural blend shapes. Our framework produces pose-dependent displacements that result in high quality deformations, especially in the joint regions. 
}
\label{fig:teaser}
\end{teaserfigure}

\maketitle

\section{Introduction}
Animating a 3D character from motion capture (mocap) data is a complex, arduous skill that animators spend years attempting to master~\cite{o2018rig}. Given a new character, a typical scenario involves manually creating a suitable character \textit{rig} that is bound to the input geometry via \textit{skinning} weights. 
Careful consideration should be given to the skeleton hierarchy of the designed character rig {so as to} correspond to the skeletal structure used in mocap data. Additionally, different poses, \emph{e.g.,} bending the elbow vs.\ extending it, require a tailored \textit{pose-specific} corrective deformation, especially in the joint regions.

A character can be articulated by applying joint rotations (obtained \emph{e.g.}\ from motion capture data) to a skeletal deformation model, typically linear blend skinning (LBS) \cite{magnenat1988joint} or dual quaternion skinning (DQS) \cite{kavan2007skinning}. 
Their simple and efficient formulation makes these methods a popular choice for animation software, games, and recently even deep learning~\cite{xu2020rignet}. However, such skinning and rigging deformation models are an oversimplification of how humans and animals move, resulting in notorious artifacts (\emph{e.g.,} elbow collapse). A high quality deformation can be obtained using blend shapes~\cite{lewis2000pose,weber2007context} where, for example, the blending coefficients are \textit{conditioned} on joint rotations~\cite{loper2015smpl} to provide fine-grained control in delicate regions.
We are inspired by SMPL~\cite{loper2015smpl} (recently extended to STAR~\cite{STAR:2020}), which achieves high quality deformations using predicted blend shapes for characters with a fixed mesh connectivity. However, a limitation of SMPL in the context of rigging and skinning is that in practice different characters almost always contain different mesh connectivities.

In this work, we present a neural technique for articulating 3D characters that learns rigging, skinning, and blend shapes for inputs with \textit{arbitrary} mesh connectivity.
We purposefully design our architecture to use a \textit{prescribed} skeleton structure, enabling generating practical skeleton rigs that are compatible with mocap data and simplifying the \textit{mocap-to-deformation} process.
Our system animates characters using enveloping with the {desired} skeletal structure and pose-specific corrective deformations. It predicts skinning weights for the input mesh and computes a set of corrective, pose-dependent shapes that improve the deformation quality in joint regions, coined \textit{neural blend shapes.}

During training, the network observes deformed shapes and learns to infer the corresponding rig, skin and blend shapes using \textit{indirect supervision}, bypassing the need to provide supervised envelope or blend-shape deformation parameters. Unlike recent deep learning techniques that supervise the network with ground-truth rigging and skinning parameters \cite{liu2019neuroskinning,xu2020rignet}, our approach does not assume that the training data has a specific underlying deformation model. {Our indirect supervision enables learning an arbitrary number of blend shapes, which we use to generate fewer blend shape bases than the original training data. Our network also learns to \emph{mask} the generated blend shapes, creating compact and localized bases without the need for such masks for supervision.}

We learn deep features directly on the input mesh connectivity using MeshCNN~\cite{Hanocka2019MeshCNN}, and predict a mesh attention map to modulate the deep vertex features, giving rise to neural skinning weights. In addition, we learn deep features on the target skeleton hierarchy using a skeleton-aware network~\cite{aberman2020skeleton} to estimate the rigging parameters. We further enrich the training data by performing mesh connectivity augmentations (\emph{e.g.,} edge collapse, flip and split), which enables us to generalize to unseen mesh connectivity during inference.

This work is the first deep learning based method for automatic enveloping combined with pose-dependent blend shapes for a skin mesh with arbitrary connectivity. Our neural blend shapes can generate high quality mesh deformations and avoid the notorious artifacts of LBS-based systems~\cite{baran2007automatic,liu2019neuroskinning,xu2020rignet}.
Our framework conforms to popular skeletal animation models, enabling plug-and-play of our output in standard animation software and game engines. We demonstrate the performance of our method on a variety of examples, including unseen, unrigged characters built by 3D artists.

\section{Related Work}
\label{sec:relatedwork}
\subsection{Mesh deformation of articulated shapes}
Deforming a mesh based on a skeletal deformation is a fundamental problem in computer graphics. One of the earliest and most widely used skinning techniques is linear blend skinning (LBS)~\cite{magnenat1988joint}. This method computes the deformation of the mesh as a weighted sum of the character's bone transformations, where the skinning weights can be computed manually or automatically. The simple formula of LBS allows fast evaluation and can be easily parallelized to fully utilize modern GPUs' high performance, making the method an essential technique for real-time applications, such as games.
\begin{figure*}
	\centering
	\includegraphics[width=\linewidth]{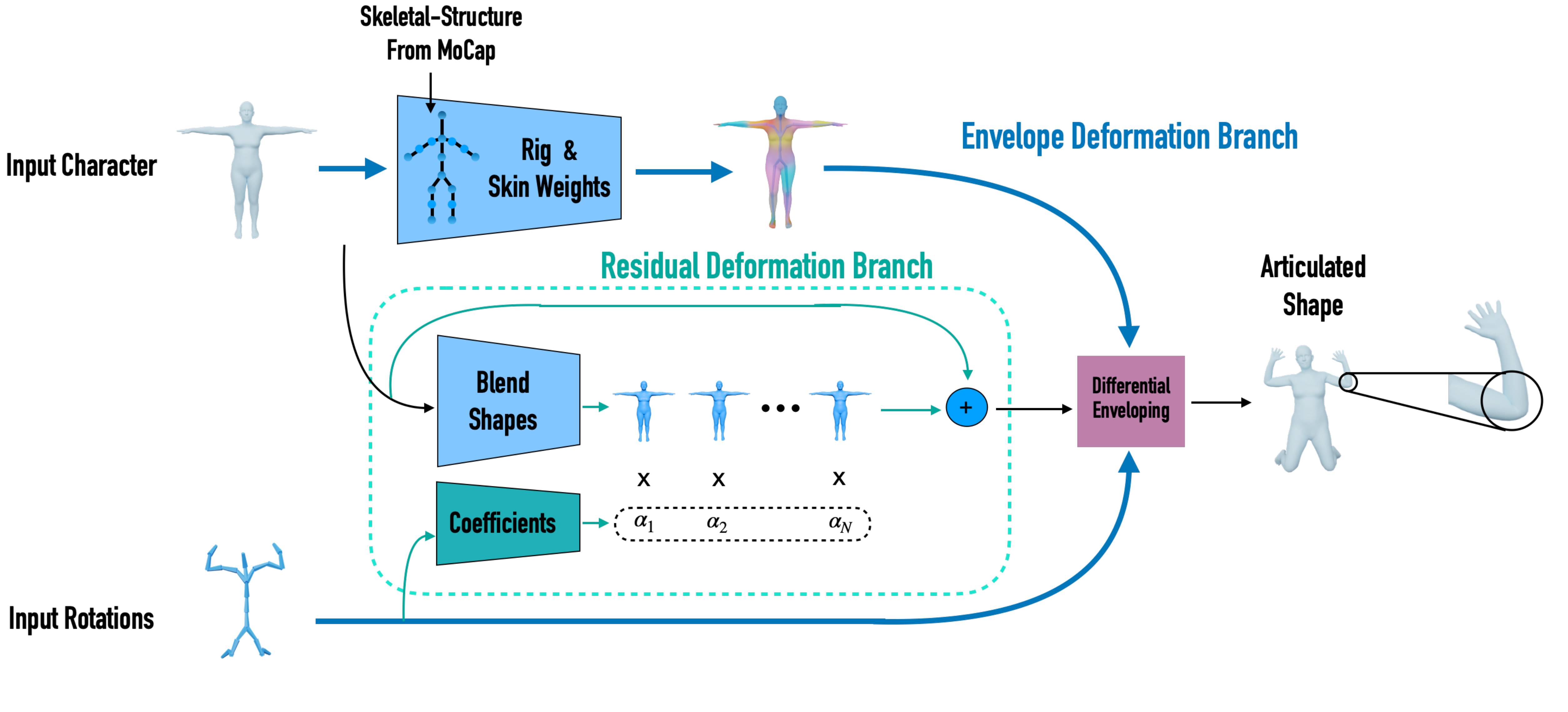} 
	\caption{Method overview. Starting with a character model in T-pose and the joint rotations on the desired skeleton hierarchy, our envelope branch predicts the corresponding skinning and rigging parameters and deforms the input character using a differential enveloping. In parallel, the residual deformation branch uses the input mesh to predict $N$ blend shapes and uses the joint rotations to predict the corresponding blending coefficients $\alpha_i$. The blend shapes add a residual deformation that is conditioned on the input joint rotations, resulting in high-quality pose-dependent deformations. Note that our system handles input characters with arbitrary mesh connectivity and does not use joint positions as input.}
	\label{fig:high_level_arc}
\end{figure*}
Despite this success, linear blend skinning suffers from known artifacts, such as \emph{elbow collapse} and \emph{candy wrapper}. Existing works develop improved techniques to overcome such shortcomings, such as dual quaternion skinning~\cite{hejl2004hardware, kavan2007skinning, le2016real}, spherical based skinning~\cite{kavan2005spherical}, multi-linear techniques~\cite{wang2002multi,merry2006animation}, and approaches with additional deformers or cages~\cite{ju2005cage,joshi2007harmonic,jacobson2011bounded,kavan2012elastic,mukai_efficient_2016,yifan2020neuralcage}. 

While LBS and similar approaches offer efficient run-time performance, they require additional corrective deformations to express details, such as wrinkles and muscle bulges, and alleviate deformation artifacts. Example-based methods provide users with more control of the deformation behavior, see \emph{e.g.}~\cite{sloan2001shape,lewis2000pose,weber2007context,frohlich2011example,loper2015smpl}.
Many such methods encode example deformations into a set of deformation bases (\emph{i.e.} blend shapes) and compute the mesh deformation as a linear combination of these blend shapes and blending coefficients. The blend shapes can be represented as vertex displacements~\cite{lewis2000pose,weber2007context}, statistical models~\cite{kry_eigenskin_2002,loper2015smpl}, or a compact sparse format~\cite{seo2011compression}.
In the standard animation pipeline, animators manually adjust blending coefficients to deform the mesh, whereas an RBF-based blend space~\cite{lewis2000pose,sloan2001shape} can facilitate this tedious process. Bone transformations and other high-level controls can be converted into blending coefficients by solving constrained geometric optimization problems~\cite{sumner2005mesh,weber2007context,frohlich2011example}. Blending coefficients can also be updated in a dynamic simulation to generate secondary effects~\cite{Hahn2012,Zhang2020}.

Recent research explores neural network based approaches to improve traditional skinning methods. Bailey et al.~\shortcite{bailey2018fast} approximate the deformation of a complex production rig using neural networks, which reduces the execution cost and allows film-quality deformation in real-time applications. Later research extends this idea to more complicated facial rigs~\cite{Bailey2020face,song2020accurate}. Neural networks can also be trained to convert high-level user control into rig parameters~\cite{Bailey2020face} to enable user-friendly editing of mesh deformation. 
Li et al.~\shortcite{li2020densegats} train a graph neural network (GNN) to apply corrective displacements to linear deformations and create nonlinear effects. While the model generates high quality mesh deformation, repeatedly evaluating a deep neural network at runtime can be expensive.
Our method is also a neural skinning technique. We employ {an envelope} skin deformation model and train a novel neural representation of blend shapes. Once generated for an input mesh, our lightweight neural blend shapes can be evaluated efficiently at runtime to achieve high quality mesh deformation.

\subsection{Automatic skinning and rigging}

Automatically rigging a skin mesh and creating ready-to-animate models has been a long-standing challenge in computer graphics.
In the pioneering work \emph{Pinocchio}, Baran and Popovi{\'c}~\shortcite{baran2007automatic} propose a template-based method that automatically fits a user-provided skeleton to a target mesh and creates an animation-ready rig.
However, this method does not generate blend shapes, and the resulting deformation can contain notorious LBS artifacts.
Miller et al.~\shortcite{miller2010frankenrigs} later demonstrate a system that automatically rigs an input mesh by combining parts from a number of templates. 
Skeleton extraction can be also achieved by analyzing the geometric features of the input mesh~\cite{Au2008,Cao2010a,bharaj_automatically_2012}. These methods are applicable to a large range of shapes, but often lack  precise control of the topology of the output skeleton.

Automatic computation of skinning weights is a complementary problem to skeleton extraction. Previous research exploits methods based on projections and heat diffusion~\cite{baran2007automatic,wareham_bone_2008}, bounded biharmonic energy~\cite{jacobson2011bounded}, geodesic voxel binding~\cite{dionne_geodesic_2013}, and physics-inspired approaches~\cite{kavan2012elastic}. 
Data-driven methods can utilize spatial coherence between examples and compute high quality skeleton and skinning weights by fitting the examples to skinning models like LBS~\cite{James2005,schaefer2007example,de2008automatic,hasler2010learning,le2014robust}. 
They can further extract blend shapes from the examples \cite{lewis2000pose,kry_eigenskin_2002,loper2015smpl}.
However, these methods require a set of example deformations of the same mesh as input, which can be difficult to obtain in practice, and they do not necessarily produce art-directable skeleton structure.

Several recent works take advantage of the power of deep neural networks to achieve high quality rigging. 
\emph{NeuroSkinning}~\cite{liu2019neuroskinning} is a GNN-based network designed to predict skinning weights. It is trained with supervised learning on a skinning dataset created by professional artists. While this model can predict high quality skinning weights, it {requires creating a suitable rigged skeleton with carefully placed joints}. 
\emph{RigNet}~\cite{xu2020rignet} is another GNN-based model for automatic rigging and skinning. The network is trained to apply mesh contraction to the input mesh and utilizes an attention-based clustering module to detect joints. The method allows users to guide the skeleton extraction with a tunable level-of-detail parameter, but there is no direct control over the topology of the generated skeleton.
Moreover, the output of this system is an LBS-based rig without blend shapes that  suffers from the standard LBS artifacts.
In contrast to these techniques, our method automatically computes rigging, skinning, and blend shapes for an input mesh. Notably, our method does not assume that the training data has a specific underlying deformation model, and our indirectly supervised training does not require ground-truth rigging and skinning parameters.

\section{Overview}
Our goal is to animate a newly designed character using available mocap data and {incorporate} high quality pose dependent deformations. In this problem formulation, it is desirable to {pre-define} the skeletal structure of the character, \emph{i.e.,} the bone hierarchy and joint adjacency, to be equivalent to the mocap data skeletal structure.

Given a new unseen character, we use a deep neural network to generate the parameters that enable articulation with pose-specific corrective deformations. The structure of our network is inspired by the classical animation pipeline and outputs three main components: rigging, skinning, and blend shapes. The number of degrees of freedom and the hierarchy of the underlying skeletal structure are pre-defined and \textit{embedded} in the network, ensuring mocap compatibility for animating the character. 

Our framework contains two main branches: (i) an envelope deformation branch that learns pose-invariant parameters (\emph{i.e.,} rigging and skinning), and a (ii) residual deformation branch that learns pose-dependent residual displacements. The learned skeleton \textit{rig} is bound to the input geometry using estimated \textit{skinning} weights. When combined with joint rotations, this defines an envelope deformation that is capable of articulating the shape.
Our neural blend shapes are inspired by SMPL~\cite{loper2015smpl}, which proposed a comprehensive model for creating exceptional deformation quality using blend shapes represented as bases of additive displacements to the input character in rest pose. 

Our network is trained on characters with the same articulation structure (\emph{i.e.,} bipeds), but which may have different underlying deformation models. Thus, no ground truth is provided to the network-estimated rigging, skinning and blend shapes. We use indirect supervision, namely, instead of directly supervising the deformation parameters (rigging, skinning, and blend shapes) our network infers them by observing how articulated vertex positions are controlled by a set of joint rotations. As a result, the network learns to represent the articulation of every input character using the pre-defined envelope model, regardless of the underlying deformation model used during training. 
{Our indirect supervision enables learning an arbitrary number of blend shapes, which we use to generate a smaller amount of blend shapes than the original training data, as well as a learned mask on the generated blend shapes. The network predicts compact and localized blend shapes that are pose-dependent \textit{by construction}, without the need for such blend shapes as supervision.}

Throughout the next sections we use the following notations:
$\bbv$ and  $\bbf$ denote the vertex positions and the connectivity of the input mesh, respectively.  $\bbw$ is the output skinning weight matrix, $\bbo$ is a hierarchical set of offsets that represent the output skeleton and $\{\bbb_i\}_{i=1}^N$ is a set of $N$ residual shapes that represent the blend shapes, interpolated with scalar coefficients $\{\alpha_i\}_{i=1}^N$. 

\section{Method}
Below we describe our neural articulation framework, which consists of two main branches:  an envelope deformation branch for rigging and skinning, and a residual pose-dependent deformation branch that enables predicting high-quality deformations.

\subsection{Envelope deformation branch}
\begin{figure}
    \centering
    \includegraphics[width=\columnwidth]{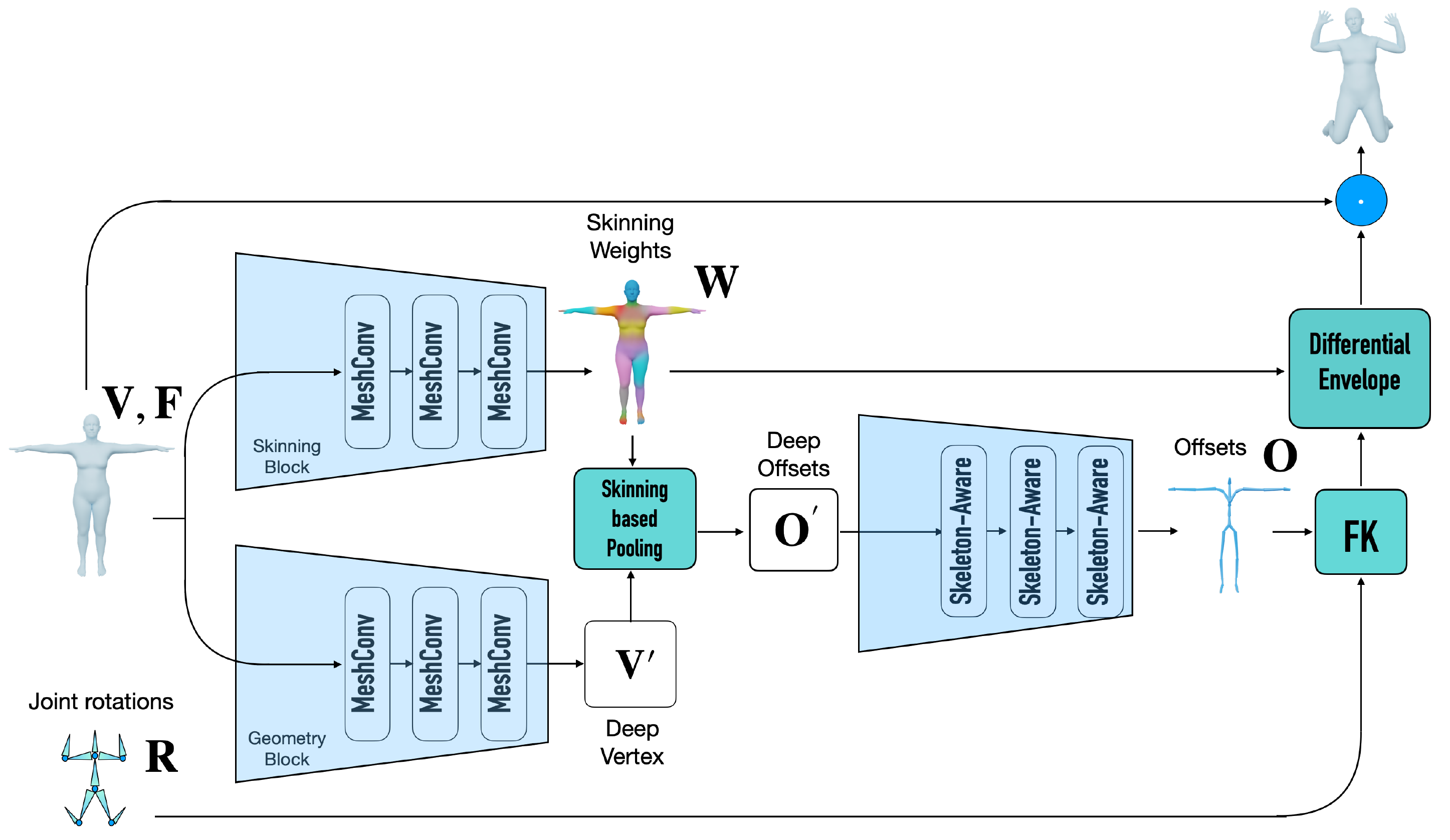}
    \caption{The envelope deformation branch. Given a mesh in T-pose ($\bbv,\bbf$) and joint rotations ($\bbr$), our network infers the skinning ($\bbw$) and rigging ($\bbo$) parameters {via indirect supervision} by 
    observing the articulated character vertex positions.}
    \label{fig:coarse_branch}
\end{figure}

Our envelope deformation branch, illustrated in Figure~\ref{fig:coarse_branch}, follows the typical animation workflow of rigging and skinning. The detailed architecture is provided in the Appendix.

Starting with an input character represented by a triangle mesh with vertices $\bbv\in \R^{V\times 3}$ and faces $\bbf$, our envelope deformation network predicts skeletal offsets (the offsets between each joint to its parent) $\bbo\in\R^{J\times 3}$ with a prescribed skeleton hierarchy that contains $J$ joints, and a skinning weight matrix $\bbw\in\R^{V\times J}$.

Since our network is not directly supervised by skinning matrices, our envelope branch can obtain better deformation quality than standard LBS.

Our network learns to fit a rig  of a pre-defined skeletal topology to any input character using indirect supervision. During training, the network is only supervised by the vertex positions of the articulated characters and the corresponding joint rotations. The network learns the relationship between joint rotations and the articulated character via the estimated rigging and skinning parameters, which are embedded in the network. Thus, with a sufficiently large number of examples, the network infers accurate rigging and skinning parameters for the input character {(for more details please refer to the experiments in Section~\ref{sec:experiments})}.

\paragraph{Skinning}
To produce skinning weights, we incorporate a series of mesh convolution blocks using the MeshCNN operators of Hanocka et al.~ \shortcite{Hanocka2019MeshCNN}. Their work demonstrates that mesh convolutions are particularly useful for classification of surface parts (\emph{i.e.,} segmentation), which is similar in spirit to skinning weights. However, our input features are different from MeshCNN. For each edge, we calculate the average positions of its two adjacent vertices.
Furthermore, we max-pool one out of five of the output channels in each hidden layer then repeat and concatenate the result along the edge axis to extend the receptive field, similar to the segmentation network presented in PointNet~\cite{qi2017pointnet}. After a forward pass, in order to predict per-vertex values, we average adjacent edge features of the corresponding vertex based on the mesh connectivity (similar to Point2Mesh~\cite{Hanocka2020point}) to get the skin matrix $\bbw$.

\paragraph{Rigging}
Given $\bbv$ and $\bbf$, our goal is to learn the rigging parameters $\bbo\in\R^{J\times 3}$ of a  specific skeleton hierarchy that consists of $J$ offsets.
Intuitively, each offset $\bbo_j$ of the character's rig can be inferred from its surrounding mesh vertices.
To learn a vertex representation that fits that task, we first pass the edge representation of $\bbv$ (similar to the skinning block) through several MeshCNN blocks to obtain a learned deep vertex representation  $\bbv^{'}\in\R^{V\times K}$ with $K$ channels. Then, the output skinning matrix is used to apply a \emph{skinning based pooling} on the deep vertices, which collapses the $V$ features into a set of $J$ deep offsets using the relative skinning weight via
\begin{equation}
\bbo^{'}_j=\frac{\sum_{i=1}^{V} \bbw_{ij} \bbv^{'}_i}{\sum_{i=1}^V \bbw_{ij}}\,,
\label{eq:lbs}
\end{equation}
where $\bbo^{'}_j\in\R^{K}$ represents a deep feature corresponding to the $j$th offset, and $\bbw_{ij}$ is the skin weight that ties vertex $i$ to offset $j$. This operation is similar to attention based pooling, and ensures that each offset is calculated only as a function of the vertices that are bound to it. 

Given the deep offsets $\bbo^{'}$, we predict the explicit skeleton offsets to construct the rig. Since the skeletal topology is fixed in our network, we can exploit joint connectivity, such that each offset is calculated only by its corresponding deep offset and its close neighbors. Hence, we use a block of skeleton-aware operators~\cite{aberman2020skeleton} to predict the explicit offset $\bbo\in\R^{J\times 3}$.

\paragraph{Envelope training}
In order to learn skinning and rigging parameters that are not provided during training, in each iteration we inject a pose described by local joint rotations $\bbr = \{\bbr_i\}$ where $\bbr_i\in\R^{3\times 3}$, which guides the deformation of the input character along with the predicted rig and skin. We use two steps to convert the local joint rotations and offsets to a global affine  per-joint transformation $\bbt_i \in \R^{4\times4}$, which can be applied to the input vertices. First, for each joint we accumulate the local affine transformations $\{\bbr_i, \bbo_i\}$ along its kinematic chain (starting from the root) through a forward kinematics layer. Then, we apply a differential linear blend skinning (LBS) layer that calculates a per-vertex global transformation based on the skinning matrix via
\begin{equation}
\bbt_{\bbr_j}=\sum_{i} \bbw_{ji} \bbt_i.
\label{eq:lbs1}
\end{equation}
Once the transformation is calculated, the per-vertex affine transformation $\bbt_{\bbr}=\{\bbt_{\bbr_j}\}$ is applied to the input character:
\begin{equation}
\tilde{\bbv}_{\bbr}=\bbt_{\bbr}\odot \bbv,
\label{eq:lbs2}
\end{equation}
where $\odot$ denotes the per-vertex operation of the global affine transformations $\bbt_{\bbr}$ on the input vertices.
An $\ell_2$-loss is applied to the difference between the reconstructed vertex positions to the ground-truth articulation $\bbv_{\bbr}$:
\begin{equation}
\Loss_{\text{v}} = \Vert \tilde{\bbv}_{\bbr}-\bbv_{\bbr} \Vert^2.
\label{eq:loss}
\end{equation}

\subsection{Residual deformation branch}
\begin{figure}
    \centering
    \includegraphics[width=\columnwidth]{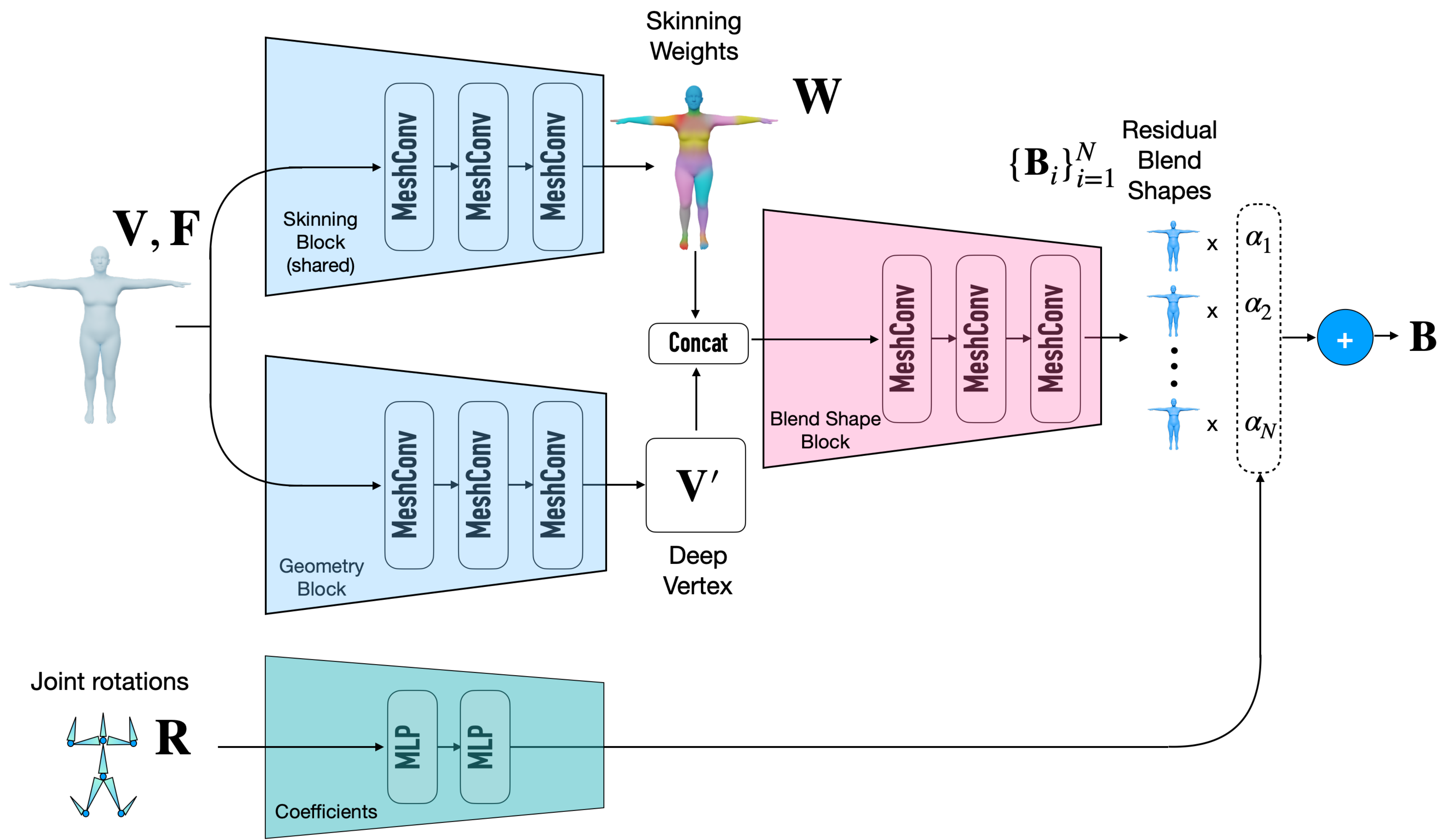}
    \caption{The residual deformation branch predicts blend shapes and blending coefficients based on the input mesh connectivity and input joint rotations. The network learns to estimate blend shapes for arbitrary mesh connectivities and blending coefficients that are conditioned on the joint rotations.}
    \label{fig:fine_branch}
\end{figure}
\begin{figure}
	\centering
	\includegraphics[width=\linewidth]{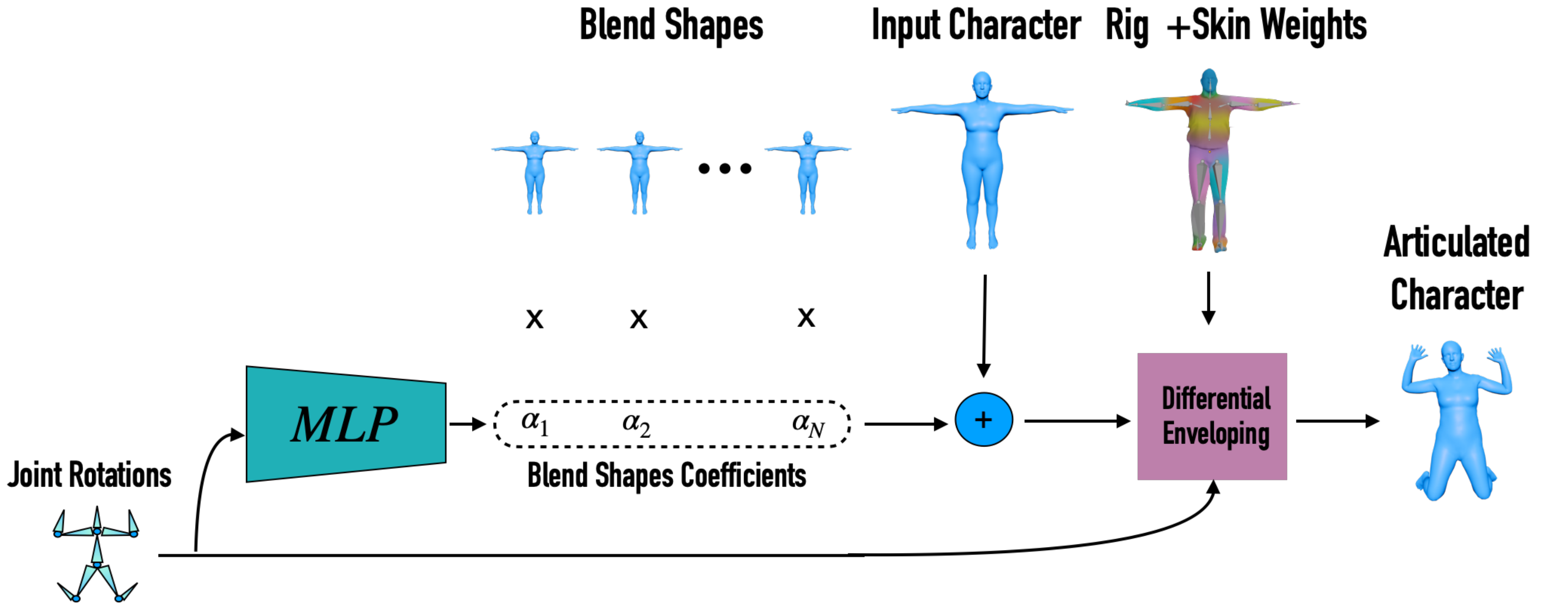} 
	\caption{Inference time. Given a new character, we can extract skin, rig, and blend shapes within a single forward pass of our network. To animate the character we need only a small network that calculates pose dependent blend shape coefficients. Conforming to standard skeletal animation models enables direct plug-and-play application in standard animation software.}
	\label{fig:inference}
\end{figure}

Our residual deformation branch is inspired by the concept of blend shapes, and predicts a set of fixed residual shapes that are interpolated by pose-dependent coefficients and added to the input character to improve the deformation quality (illustration in Figure~\ref{fig:fine_branch}). In our case, both the shapes and their coefficients are learned by a neural network. During inference we feed the input character to the network once to receive the residual blend shapes, whereas the joint rotations for every frame are needed to animate the pose-dependent deformations for the character in real-time, as illustrated in Figure~\ref{fig:inference}.

\paragraph{Residual blend shapes}
\begin{figure*}[h]
    \centering
    \includegraphics[width=\linewidth]{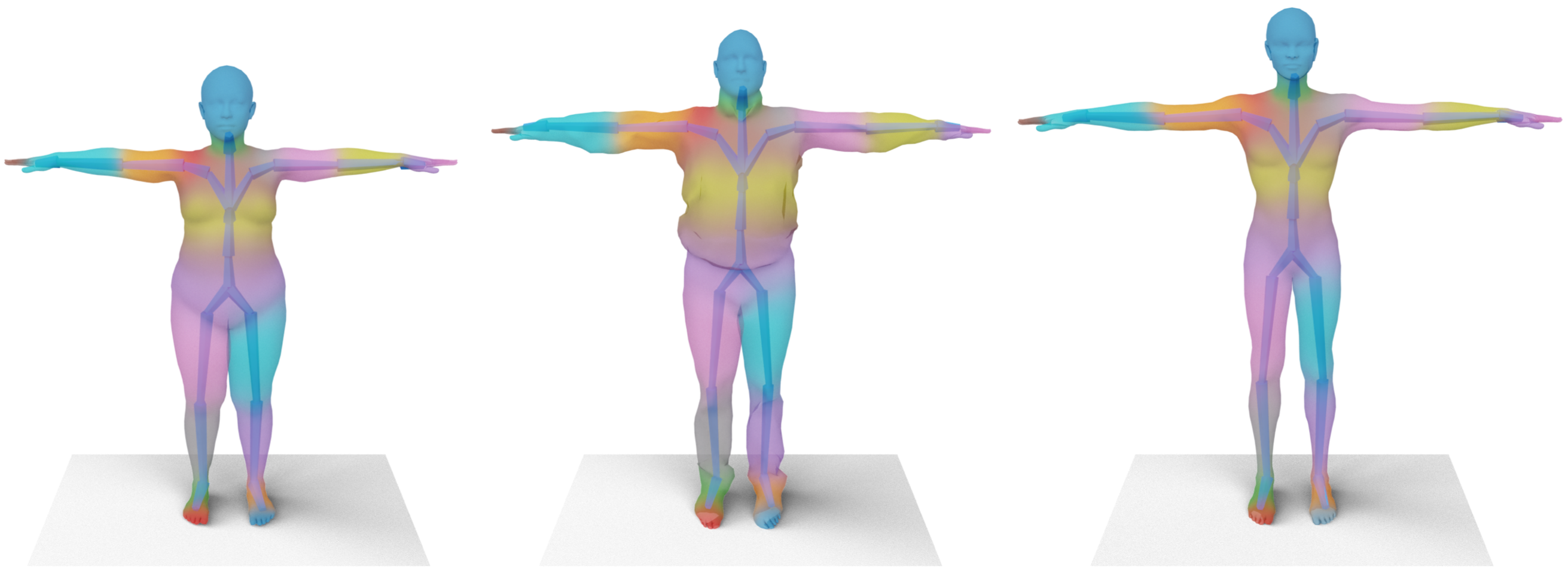}
    \caption{Visualization of the network predicted rigging and skinning weights for various unseen characters. From left to right we see random test characters from each of the three test set groups: SMPL dataset, garment dataset, and characters handcrafted by 3D animator.
    }
    \label{fig:rig_skin_result}
\end{figure*}
Given the input vertex positions $\bbv$ and connectivity $\bbf$, the residual branch starts with the skinning and geometry blocks with {fixed} weights that were  pretrained in the envelope deformation branch. 
Then the output skinning $\bbw$ is concatenated to the deep vertices $\bbv^{'}$ along the channel dimension ($\{\bbv^{'},\bbw\}\in\R^{V\times({K}+J)}$) and the result is fed into the network.
The combination of deep vertex and skin provides the blend shapes network with rich information about the vertices and their relationship to the skeleton, which is essential for the generation of the blend shapes.
Similar to the envelope branch, we use the edge feature representations of these three components, which are passed through a block of mesh convolutions, resulting in a set of $N$ residual shapes $\{\bbb_i\}_{i=1}^N, \ \bbb_i\in\R^{V\times 3}$.
In parallel, we feed a small neural network that contains $J$ MLP blocks, where each is conditioned by a single joint rotation, and output a series of pose dependent coefficients $\{\alpha_{ij}\}_{i=1}^{N}$ per joint $j$.
These coefficients are used to interpolate between the residual shapes that are summed up and added to the input vertices:
\begin{equation}
\tilde{\bbv} = \bbv + \sum_{j=1}^J \sum_{i=1}^N \alpha_{ij} m_j\bbb_i,
\end{equation}
where $m_j$ is a binary mask that specifies the vertices that are associated with joint $j$. This computation is done by picking all the non-zero entries in the skinning matrix that are associated with the two bones corresponding to the joint. This operation enables us to enforce localization in the structure of the blend shape and avoid undesired deformation of vertices associated with static joints (similar to \cite{STAR:2020}).
Similar to the envelop branch, the loss is calculated as the difference between the articulated character and the corresponding ground truth using Eq.~\eqref{eq:loss}.

\section{Experiments and Evaluations}
\label{sec:experiments}
In this section we evaluate our results, compare them to other rigging, skinning, and deformation techniques, and demonstrate the effectiveness of various components in our framework through ablation study. In order to qualitatively evaluate our results to the fullest extent, please refer to the supplementary video.

\subsection{Implementation details}
Our neural articulation framework is implemented using the PyTorch library~\cite{pytorch}, and the experiments were performed on NVIDIA GeForce GTX Titan Xp GPU (12 GB) and Intel Core i7-6950X/3.0GHz, CPU (16 GB RAM). 

We train the network using a two stage course to fine approach. In the first phase, we train the envelope branch, and in the second phase, we fix the envelope network weights and train the residual branch. In this phase, we found that for our specific training dataset, we can boost the performance by providing supervision for the blend-shapes. However, since the blend-shape of SMPL and our model do not share the same properties (number of blend-shapes, number of vertices per blend shape) we extract our blend-shapes ground truth samples by optimizing the MLP network and $\{\tilde \bbb_i\}_{i=1}^N$ such that the output satisfies some high quality deformed ground truth. Then the extracted blend shapes can be used to supervise directly the generation of residual shapes $\{\bbb_i\}_{i=1}^N$. We optimize the parameters of our framework using the Adam optimizer~\cite{kingma2014adam}. We use different learning rates for the different blocks, and these are specified in Table~\ref{tab:arch}. Training our network took about 3 days.

\begin{figure*}[h]
    \centering
    \newcommand{\pll}{-1}
    \begin{overpic}[width=\linewidth]{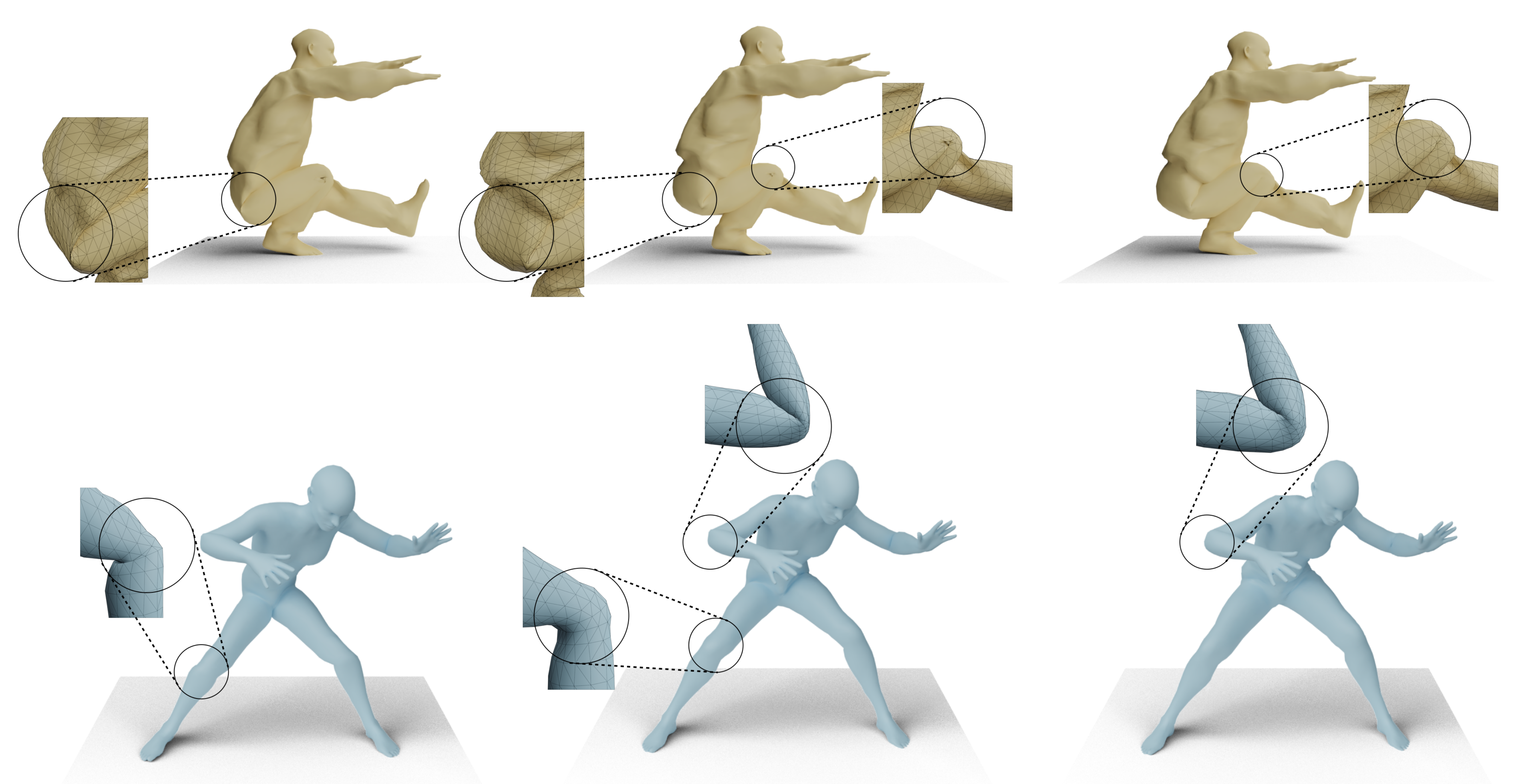}
    \put(15,  \pll){\textcolor{black}{LBS}}
    \put(48,  \pll){\textcolor{black}{Envelope}}
    \put(75,  \pll){\textcolor{black}{Envelope + Blend Shapes}}
    \end{overpic}
    \caption{{Our predicted envelope deformation produces favorable results, compared to LBS. Adding neural blend shapes to the enveloping results in corrective pose-dependent displacement, which improves the deformation quality in the joint regions.}}
    \label{fig:bs_comparison}
\end{figure*}

\subsubsection{Data}
Our network is trained on the SMPL dataset~\cite{loper2015smpl} which contains
ten \emph{shape} (pose-independent) blend shapes and 207 pose-dependent blend shapes. This enables generating a variety (\emph{e.g.,} height, weight, proportions) of different shape identities and high quality deformations using the joint rotations provided in the SMPL model. Since the SMPL shapes represent a relatively \emph{sterile} character (\emph{i.e.,} naked and hairless humans), we incorporate an additional dataset of clothed humans proposed in Multi Garment Network~\cite{bhatnagar2019mgn}. The latter contains 96 characters with SMPL mesh connectivity. We used 80 out of the 96 clothed humans for training (reserving 16 models for testing). 
To train the network, we sample joint rotations from two different distributions (one for each branch). For the envelop deformation branch, the distribution for a single joint is $U \left( \mathbb{S}^2 \right) \times \mathcal{N}\left( 0, (\pi / 6)^2 \right)$, namely, the rotation axis is uniformly sampled from a 3D sphere and the rotation angle is sampled from a normal distribution with zero mean and variance of  $(\pi / 6)^2$.
For the residual deformation branch, the rotation distribution is $U \left( \mathbb{S}^2 \right) \times U\left[0, 2\pi\right]$ to enable capturing of even larger and exaggerated deformations. In each branch we sample rotations for each joint except for the root joint. 

Our network assumes that the input character has a consistent upright and front facing orientation. Following SMPL~\cite{loper2015smpl}, the input should also be in T-pose in order to effectively learn blend shapes, which is important for obtaining high quality deformations. In addition, the vertex positions are spatially aligned during training such that the hand tips are in a fixed height (as in the SMPL model). In test time, if the character does not belong to the SMPL distribution, we translate it to the same height and scale the vertices such that the two extreme hand tip vertices have a certain distance, similar to the distance in the raw SMPL model, with no shape blend shapes.

As outlined in MeshCNN~\cite{Hanocka2019MeshCNN}, our system can handle inputs which contain boundaries, self-intersections, or disconnected components. In the case of non-manifold geometries (non-manifold mesh edges or vertices), these should simply be deleted. In general, there should be a meaningful shape surface to propagate the deep mesh features; so in the case of an extreme polygon soup some form of remeshing~\cite{meshmixer} or tetrahedralization~\cite{ftetwild} can be employed.

\paragraph{Test data} Our test set consists of three groups. The first group is five biped characters manually created by a professional animator. The second group is thirty random SMPL characters, which are sampled with different proportions and shapes than we trained on. The third group is a random subset of 16 characters from the SMPL clothes dataset that were reserved for the test set.

\paragraph{Garment augmentation}
We also use the dataset of Bhatnagar~\etal~\shortcite{bhatnagar2019mgn} to enlarge the geometric features observed by our network. Since the shapes in this dataset have the same connectivity as SMPL, we can extract the garment displacements in two steps. First, we find the closet SMPL character that matches the surface of the clothed character (via optimization on SMPL parametric space), and then we subtract the clothed character from the fitted SMPL character to get a set of displacements, which are used to augment shapes by adding garments. This creates a variety of geometric variability in the training data, especially with regard to the location of the skin w.r.t. the character bone. We show the importance of this augmentation in the ablation study.

\paragraph{Mesh connectivity augmentation}
In order to enrich the set of samples that the network is trained on and to enhance its robustness to different mesh connectivity, we augment the data samples by applying three topological operators on edges: collapse, split and flip~\cite{hoppe1996progressive,botsch2010polygon}.

\subsection{Experiments}

Our framework can generate high quality rigs with skin weights and blend shapes on various characters with arbitrary connectivity. Figure~\ref{fig:rig_skin_result} shows some of our results. The test characters shown here are from each of the three test data groups: the SMPL dataset, the garment dataset, and the characters handcrafted by a 3D animator. None of these characters are used during training. In order to visualize the skinning weights we associated each bone in the skeleton with a unique color, then the color of the vertex is calculated as weighted average of the colors of the bones bound to it in the skinning weights. 

Figure~\ref{fig:bs_comparison} demonstrates the quality of the deformations generated by our predicted skeleton rigs. The baseline deformations on the left are generated using LBS weights and the ground-truth skinning weights of each character, while the deformations generated by our envelope deformation branch and residual deformation branch are shown in the middle and on the right of the figure, respectively. 
Notably, because our envelope branch is trained on high-quality data, we can already generate mesh deformations that are better than the baseline results using LBS and the skinning weights computed by our envelope model. This improvement can be easily spotted in Figure~\ref{fig:bs_comparison}, where the volume loss around the buttocks area is significantly mitigated in the results showing in the middle.
The quality of the deformation can be further improved using the blend shapes computed by our residual deformation branch. As shown in Figure~\ref{fig:bs_comparison}, the volume of the meshes around the knee and the elbow are preserved even when the corresponding joints are transformed dramatically. 

\begin{figure}
\centering
\begin{tabular}{c}
	\includegraphics[width=\linewidth]{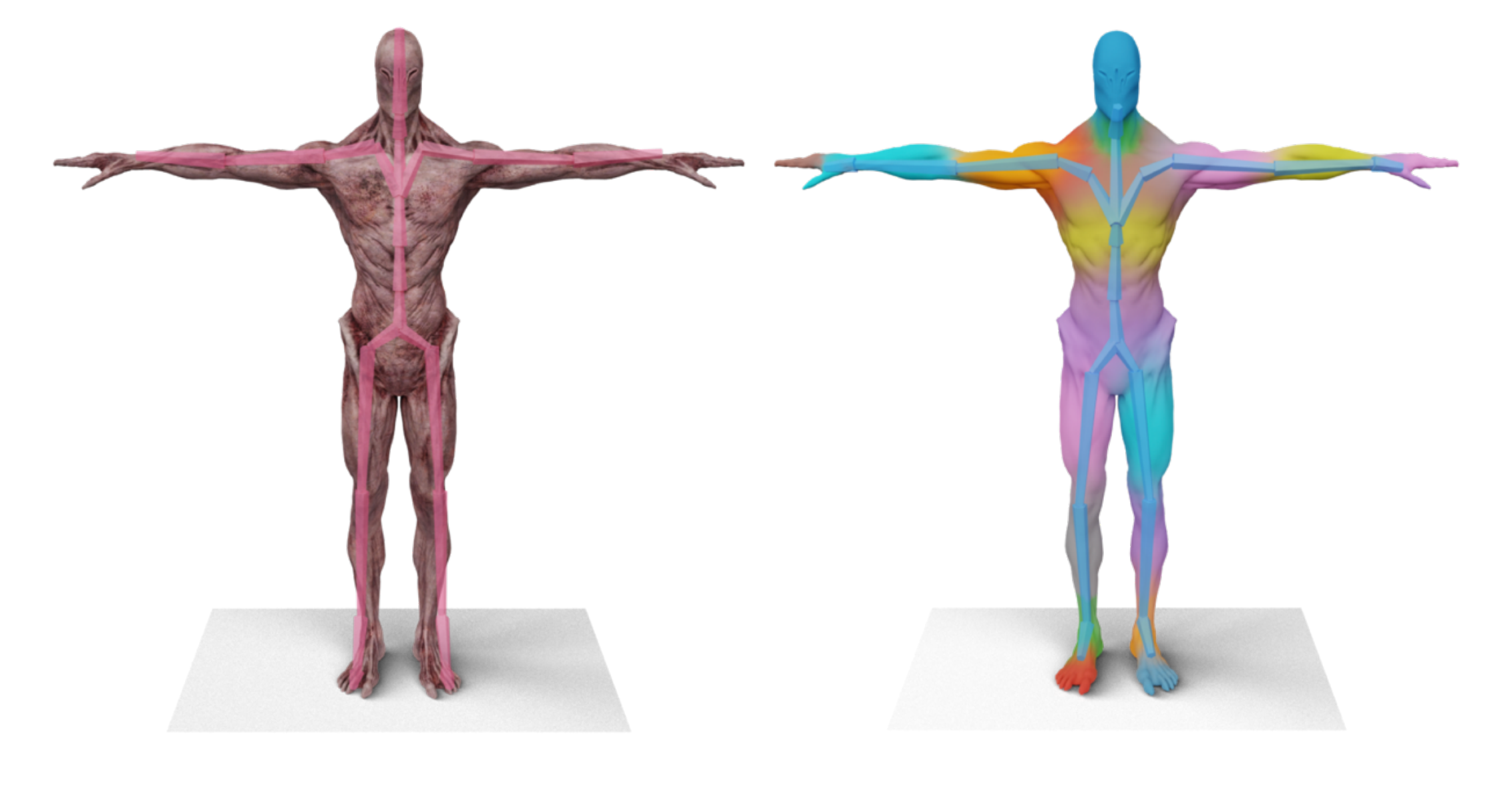} \\
	(a) \\
	\includegraphics[width=\linewidth]{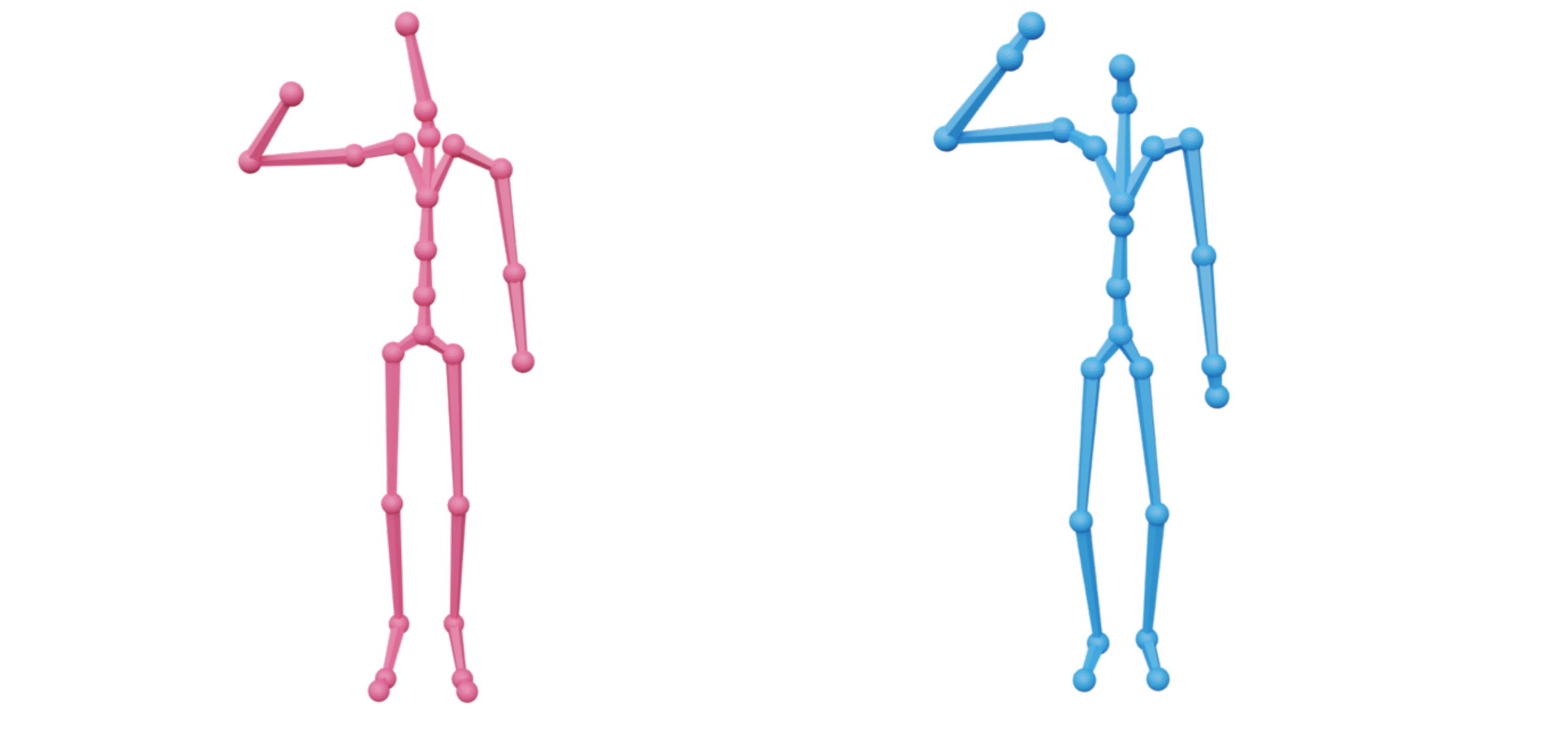} \\
	(b) \\
	\includegraphics[width=\linewidth]{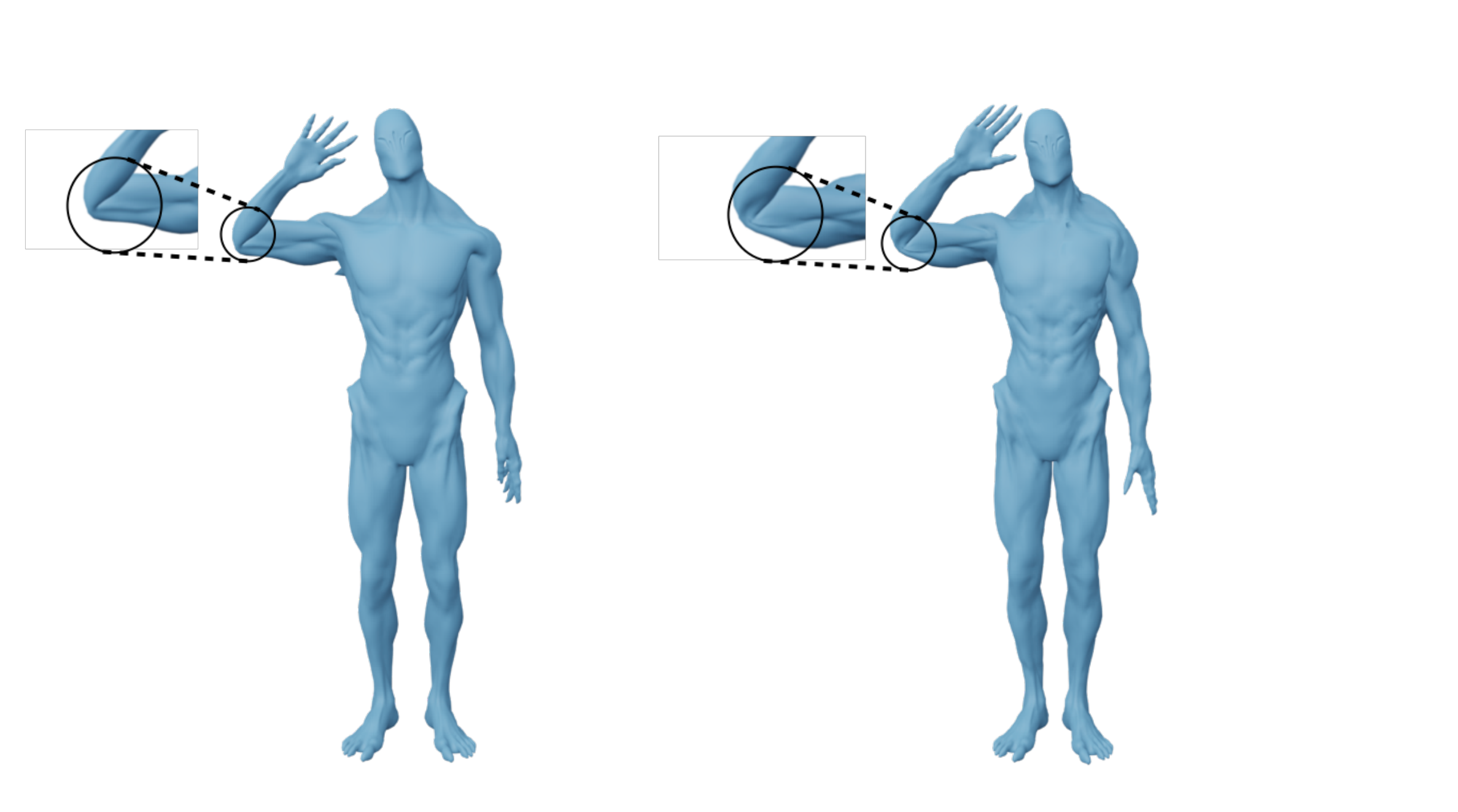} \\
	(c) \\
	Mixamo\ \ \ \ \ \ \ \ \ \ \ \ \ \ \ \ \ \ \ \ \ \ \ \ \ \ \ \ \ \ \ \ \ \ \ \ Ours
\end{tabular}
\caption{Our system generalizes to characters from other datasets (Mixamo). (a) Original Mixamo character and its skeleton (left) and our mocap-ready output skeleton with the corresponding skinning weights (right).  (b) Original Mixamo skeleton posed (left) and our mocap-ready skeleton posed (right). (c) Original Mixamo deformation (left) and our high quality deformation which is achieved by the corrective blend shape that predicts muscle bulge (right). }
\label{fig:mixamo_unified}
\end{figure}

A challenging task for neural networks is the ability to generalize and extrapolate beyond the training data, which will always contain only a \textit{subset} of the real world data we expect to encounter in test time. We validate how robust our system is to characters found \emph{in the wild}, by animating characters from the Mixamo dataset~\cite{mixamo} using our model which has not seen Mixamo characters during training. Mixamo is a particularly challenging set since the characters contain vastly different mesh connectivity, body proportions, and decorative geometries. While Mixamo already contains rigging and skinning, it does not contain the required skeletal structure for animating with a given mocap data, and there are no blend shapes or pose-dependent corrective deformations. To this end, we use our system to predict a rig with a desired skeletal-structure, enabling animating Mixamo characters using mocap motions with neural blend shapes. This result is especially remarkable, since the mesh contains three times the amount of vertices which we trained our network on. In Figure~\ref{fig:mixamo_unified}(a), we see the result of our skinning and rigging on the Mixamo character, where our predicted skeleton rig is ready to animate with the given mocap data. Our neural blend shapes creates a corrective deformation in the muscle area resulting in accurate preservation of the muscle bulging, whereas as the original Mixamo dataset does not incorporate high quality deformations as shown in Figure~\ref{fig:mixamo_unified}(c) and the supplementary video. In addition, see Figure~\ref{fig:mixamo_unified}(b), where the original Mixamo character rig is not compatible with mocap skeleton structure, whereas our network predicts a mocap-ready skeleton on the same character.

\begin{figure}[h]
    \centering
	\includegraphics[width=\columnwidth]{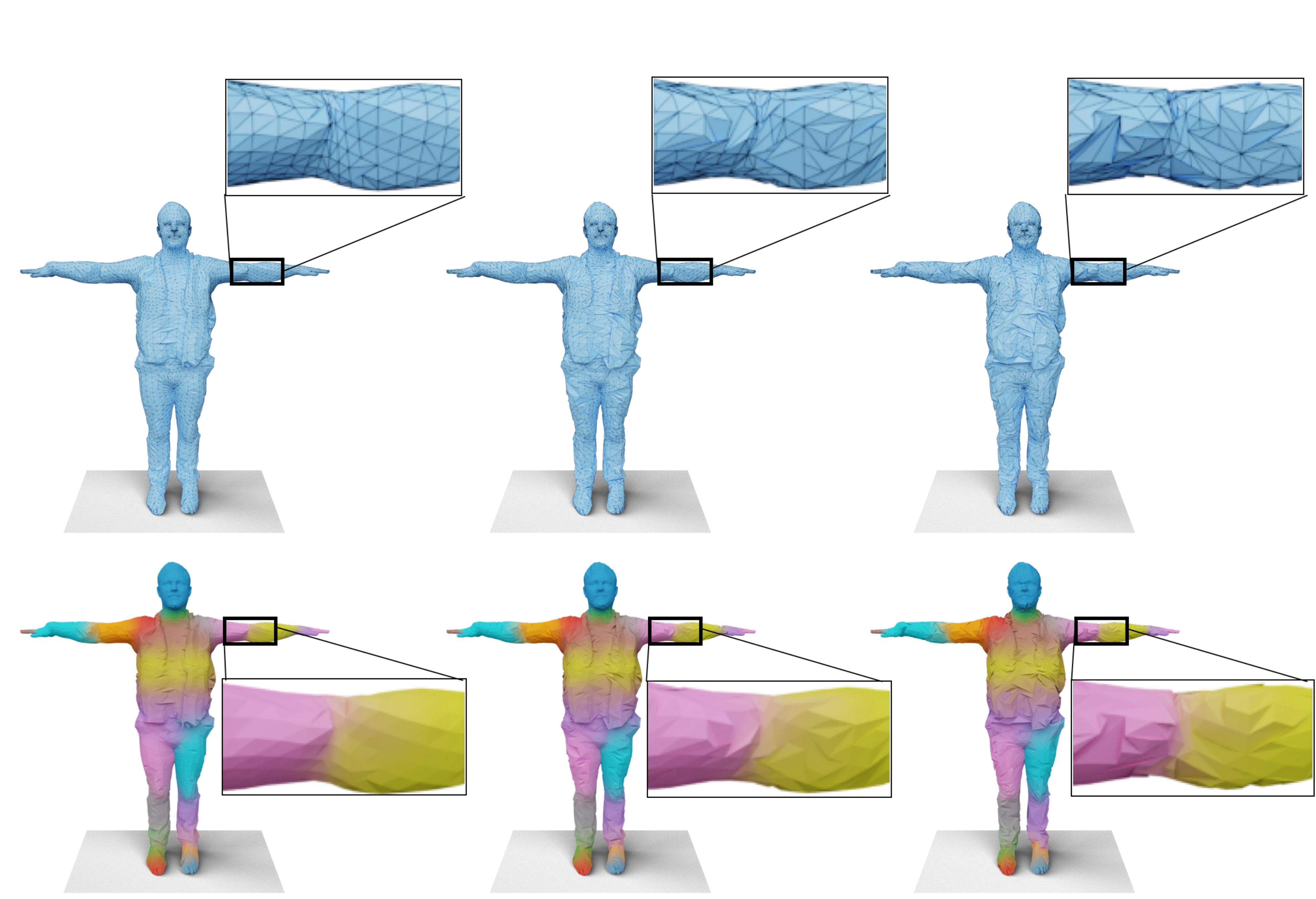}
    \caption{Robustness to variations in connectivity. Given the same input character, we perform connectivity augmentations and observe that the corresponding skinning weights are stable.}
    \label{fig:connectivity_robustness}
\end{figure}

\paragraph{Connectivity robustness}
In this experiment, we demonstrate that training with mesh connectivity augmentations (edge flip, split, collapse) results in a system that is robust to changes in the input mesh connectivity. Given the same input character, we perform a significant amount of connectivity augmentations to achieve variations of the mesh input. We observe that the network estimated skinning and rigging parameters remain stable, which can be seen in Figure~\ref{fig:connectivity_robustness}. We also show that the corresponding enveloping produces stable deformations for such input connectivity variations, which can be seen in the supplementary video. 

\begin{figure}[h]
    \centering
\includegraphics[width=\linewidth]{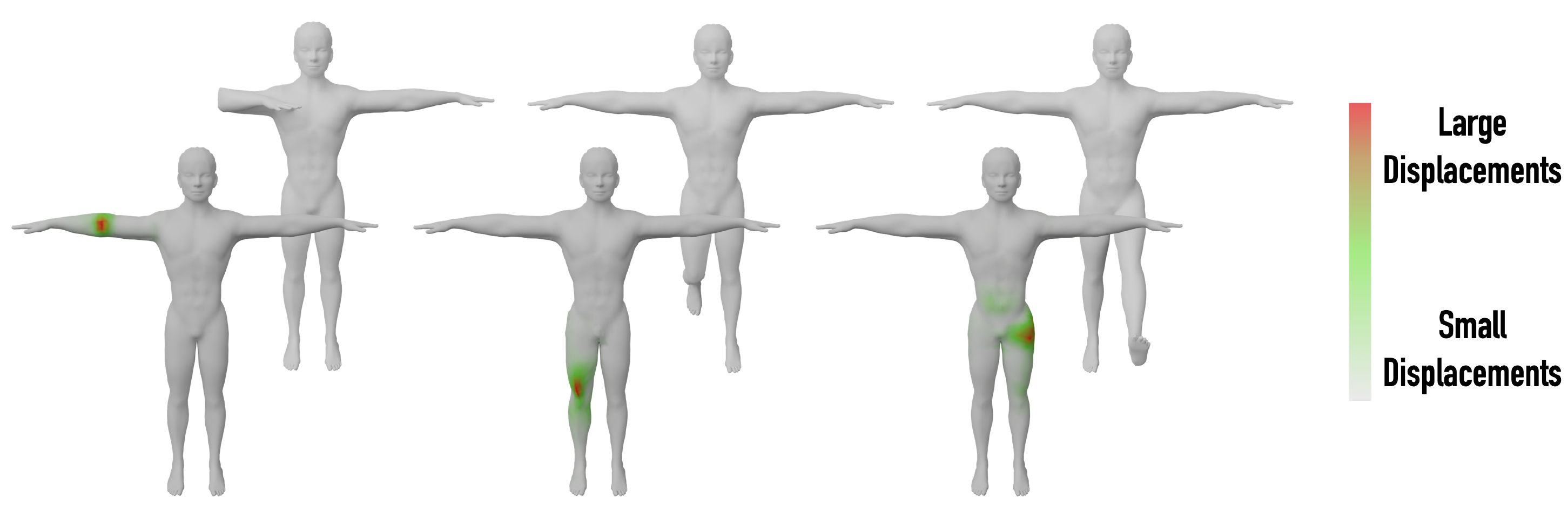}
    \caption{Neural blend shape corrective displacement visualization. Top: the same shape is shown in three different poses, bottom: corresponding blend shape displacement (visualized with a heat map). Observe that a rotation in the elbow, knee, and hip joints resulted in corrective deformations in the corresponding regions.}
    \label{fig:blend_shapes_visualization}
\end{figure}

\paragraph{Neural Blend Shapes}
In this experiment we visualized the corrective effect of the learned pose dependent blend shapes. For a given character, we fed our network with 3 sets of joint rotations (equivalent to 3 poses) and visualized the output in Figure~\ref{fig:blend_shapes_visualization}. 
The top row exhibits the deformed character for each of the poses, while the bottom raw visualizes the magnitude of the resulting displacements (via color map), on top of the same character in rest pose. It can be seen that our neural blend shapes displace the vertices in regions that correspond to the bent joint (elbow, knee, and hip joints) resulting in a corrective deformation in these regions.

\begin{figure}
    \centering
\includegraphics[width=\linewidth]{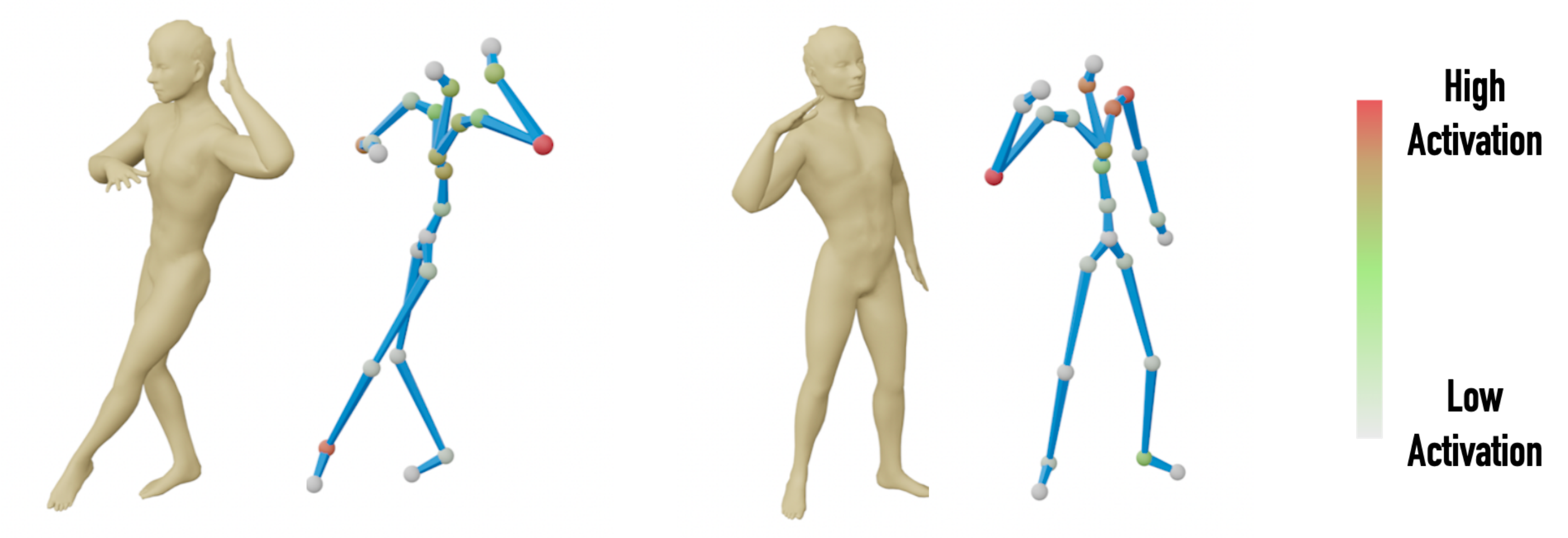}
    \caption{Neural blend shape coefficients visualization. In each pair the left shape is our output deformed character, and on the right is the corresponding posed rig. The joints are colored by the blend shape coefficient activation corresponding to the joint.}
    \label{fig:blend_shapes_coeff_visualization}
\end{figure}

In addition to the corrective displacements, we also visualize the learned blending coefficients on top of a particular motion in order to examine the active joints which can be seen in Figure~\ref{fig:blend_shapes_coeff_visualization}. For each pose, each joint is colorized according to the average values of the displacement that are associated with it based on the skinning matrix. The results demonstrate that when the joint is active (bent) the activation of the corresponding coefficients is higher. Please refer to the supplementary video to see the full animation.

\begin{figure}[h]
    \centering
    \includegraphics[width=0.7\linewidth]{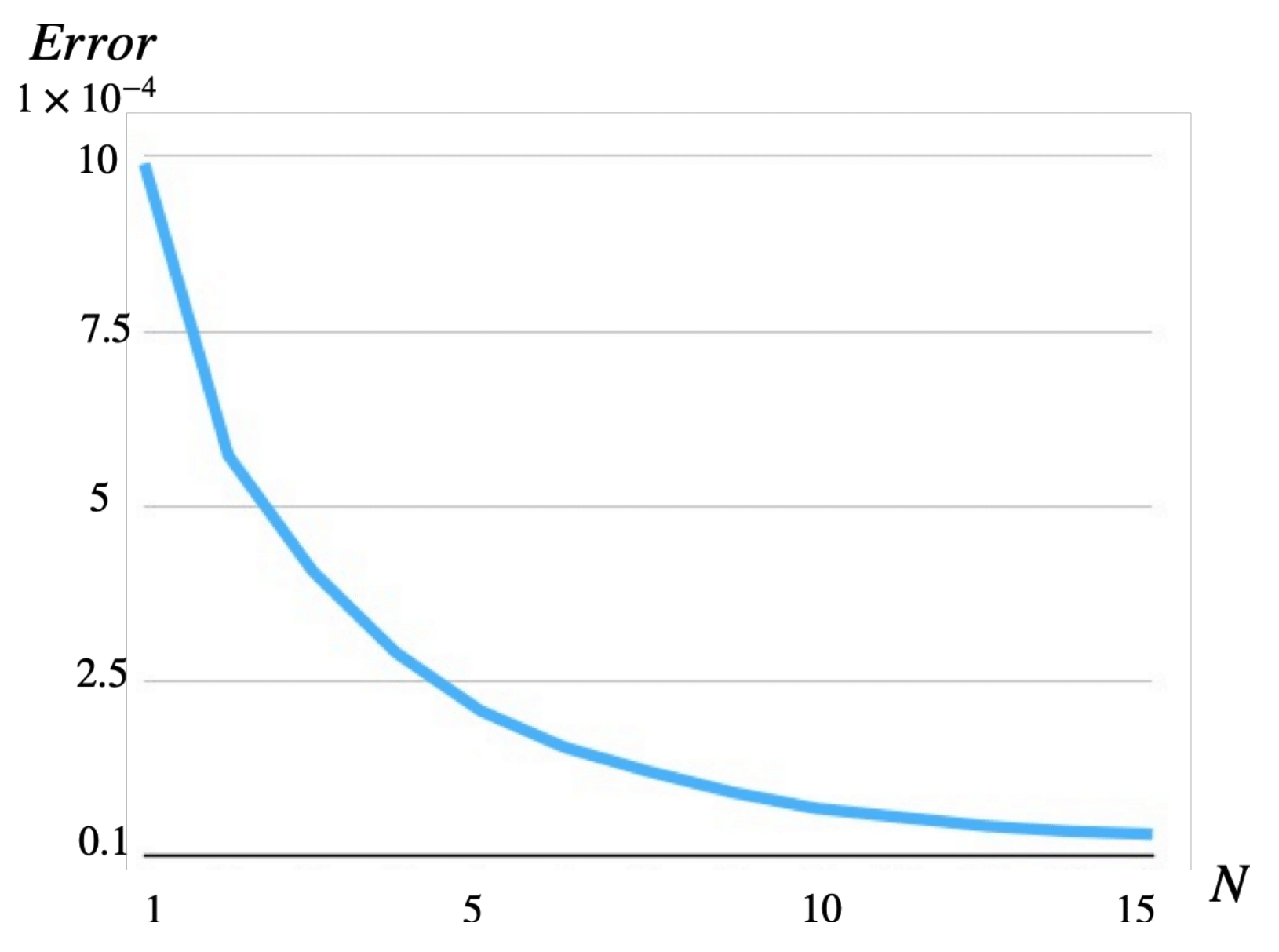}
    \caption{Number of blend shapes as a function of high quality deformation error.}
    \label{fig:blend_shape_error}
\end{figure}

In addition, we ran an experiment to evaluate the number of blend shapes needed to obtain high quality pose dependent deformations. We trained the network with a varying number of blend shapes $N$ and found that $9$ blend shapes was enough to obtain high quality deformations, both quantitatively and qualitatively. See the quantitative results in Figure~\ref{fig:blend_shape_error}.

\subsection{Evaluation}
In this section we compare our method to state-of-the-art skinning and rigging methods as well as off-the-shelf tools.

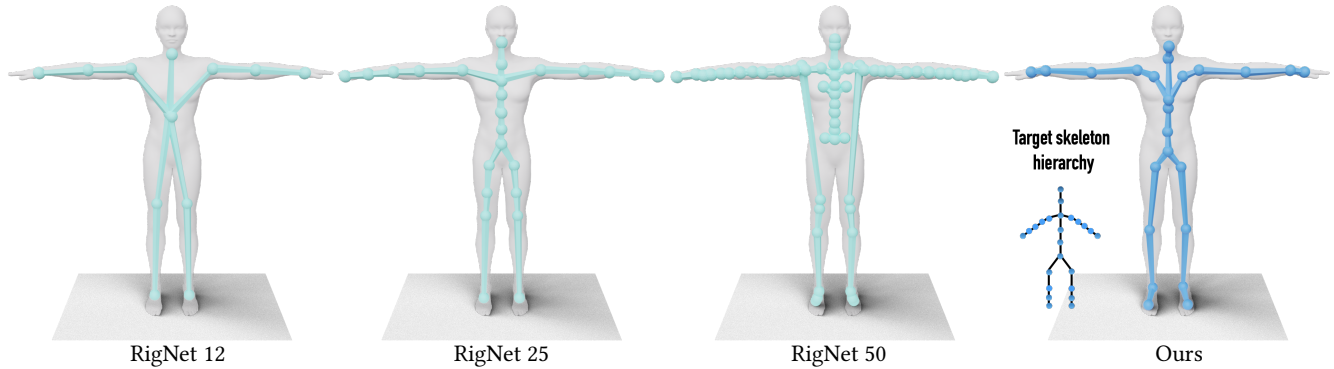
\begin{figure*}[h]
    \centering
    \newcommand{\pll}{-1}
    \newcommand{\nb}{50}
    \begin{overpic}[width=\linewidth]{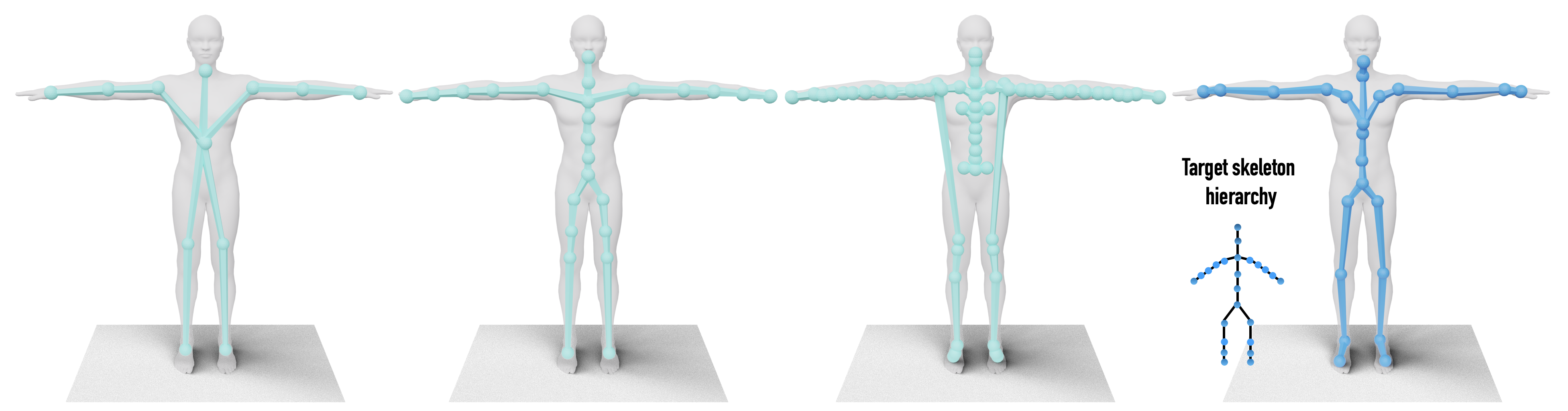}
    \put(10,  \pll){\textcolor{black}{RigNet {12}}}
    \put(34,  \pll){\textcolor{black}{RigNet {25}}}
    \put(59,  \pll){\textcolor{black}{RigNet {50}}}
    \put(86,  \pll){\textcolor{black}{Ours}}
    \end{overpic}
    \caption{Predicted skeleton rig on a character designed by a 3D artist. RigNet~\cite{xu2020rignet} only provides scalar control over the granularity of the skeleton (shown: predicted rig with 12, 25, or 50 bones corresponding to the scalar 0.015, 0.028, 0.09), but cannot control the skeleton hierarchy. We were not able to obtain the desired number 24, which is achieved by the output of our method, complying with the target skeletal hierarchy (rightmost example). Note the undesirable placement of the root node, and extraneous joints in the knee region.}
    \label{fig:rignetcmp}
\end{figure*}

\paragraph{Rigging}
We first compare to the recent work of Xu~\etal~\shortcite{xu2020rignet} which proposed a deep neural network that learns to generate arbitrary skeletal rigs and the corresponding skinning weights using supervised learning, and demonstrated impressive performance on shapes with varying skeletal hierarchies. However, their system does not enable the user to control the output skeletal hierarchy but only allows to modify the density of the predicted joints through a single scalar. Yet, animating an arbitrary skeleton rig using mocap data is not directly possible due to incompatibility in the skeletal structure. The latter requires motion retargeting between two skeletons with different, unseen, structures, which is still an open problem~\cite{gleicher1998retargetting,aberman2020skeleton}. Although the recent work of Aberman~\etal~\shortcite{aberman2020skeleton} proposed a method to retarget motion of skeletons with different hierarchical structures, it requires having datasets contain different characters with the exact same hierarchical structures for both source and target skeletons, which is incompatible with our setting.
Figure~\ref{fig:rignetcmp} shows 3 different outputs of RigNet for an input character that was designed by a 3D artist and for different input scalar values (0.015, 0.028, 0.09). It can be seen that every output rig contains different number of joints (12, 25, 50) and different skeletal hierarchies which can not be controlled directly by the user. In practice, we tried to find the scalar value that leads to 24 joints -- the number of joints in our target skeleton, and couldn't (the closet we found is 25).
In contrast, our network predicts skeleton with a fixed hierarchy which is embedded in the network, thus, can be animated with corresponding mocap data.

\begin{figure*}[h]
    \centering
    \newcommand{\pll}{-1}
    \begin{overpic}[width=\linewidth]{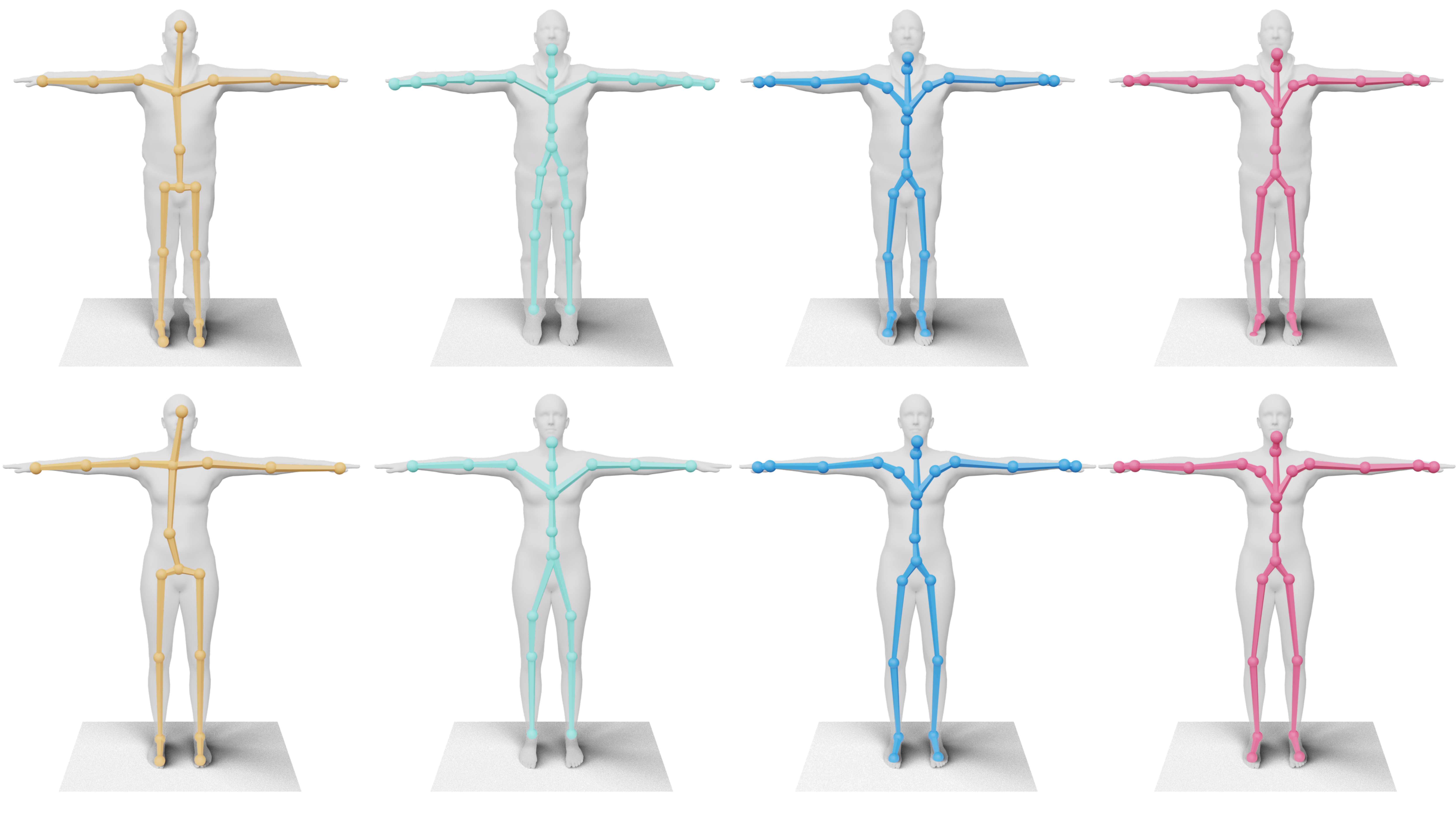}
    \put(7,  \pll){\textcolor{black}{Pinocchio~\shortcite{baran2007automatic}}}
    \put(33,  \pll){\textcolor{black}{RigNet~\shortcite{xu2020rignet}}}
    \put(61,  \pll){\textcolor{black}{Ours}}
    \put(83,  \pll){\textcolor{black}{Ground truth}}
    \end{overpic}
    \caption{Automatic rigging results on two different humanoids. Left to right: Pinocchio~\shortcite{baran2007automatic}, RigNet~\shortcite{xu2020rignet}, ours and ground truth.}
    \label{fig:rigcmp}
\end{figure*}
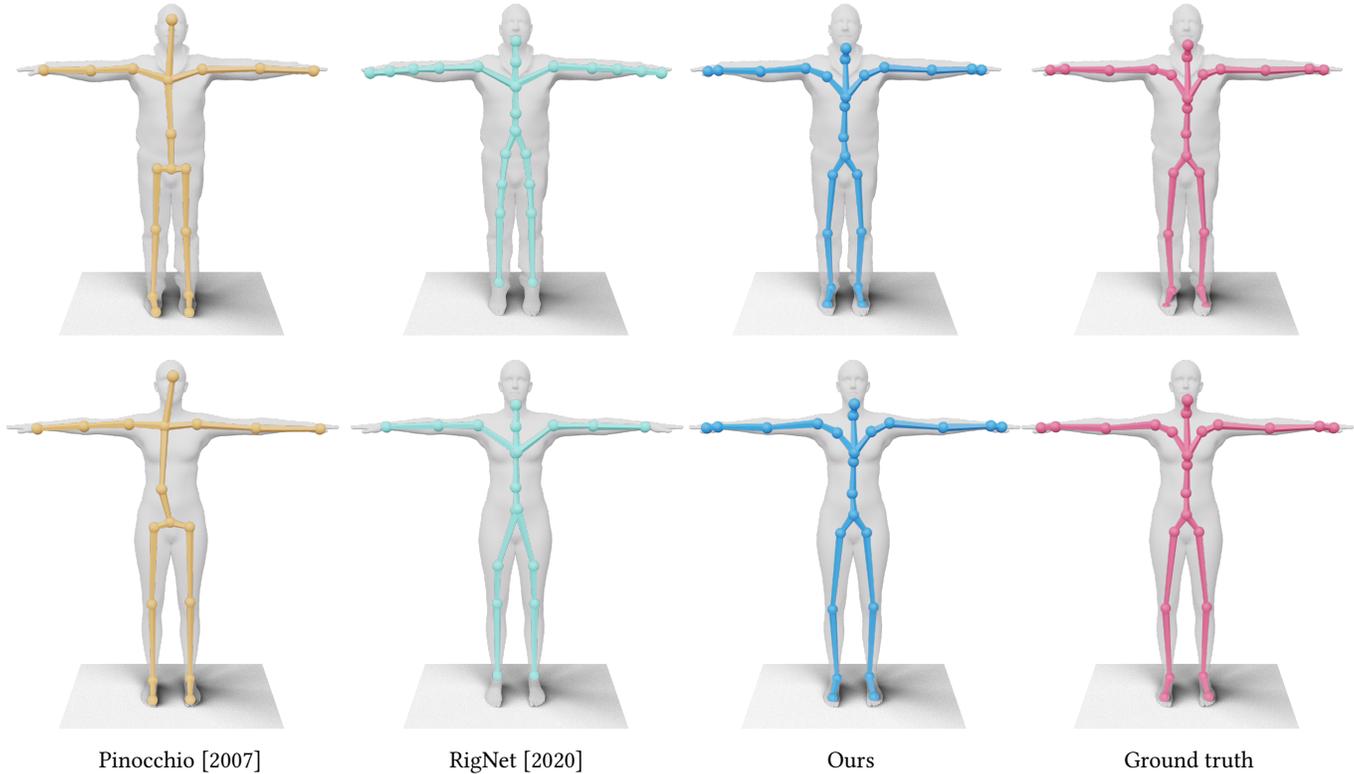

We next compare our results to the method of Baran and Popovi{\'c} \shortcite{baran2007automatic} (a.k.a Pinocchio) which fits a template skeleton for each input character. The target template is selected from a set of predefined skeletal hierarchies based on a cost function that evaluates its geometric fitting to the input shape.
Figure~\ref{fig:rigcmp} shows the comparison to the automatic rigging results of Pinocchio and RigNet for two different humanoids. It can be seen that the selected output skeletal hierarchy of Pinocchio is sparse (18 joints), which limits the granularity of deformation  that can be achieved and requires a manual specification of joint correspondence in order to animate the output with motion from mocap data.

We employ the metrics CD-J2J, CD-J2B, and CD-B2B proposed by \cite{xu2020rignet} to quantitatively evaluate the automatic rigging results. Briefly speaking, these metrics evaluates the quality of a rigging by measuring the spacial distances between the joints and bones. Ideally,
all CD-J2J, CD-J2B, and CD-B2B measures should be low for a high quality rig. We refer interested reader to the original paper for detailed definitions of these metrics.
The results are reported in Table~\ref{tab:rig_quantitative}. It can be seen that our method outperforms the other methods quantitatively. Note that the metrics enables to calculate distances between skeleton with different hierarchies.

\begin{table}
\caption{Quantitative comparison between our rigging results to the Pinocchio \cite{baran2007automatic} and RigNet \cite{xu2020rignet}.}
\begin{tabular}{l c c c}
\toprule
& \small CD-J2J &  \small CD-J2B &  \small CD-B2B \\
\midrule
\small  Pinocchio \cite{baran2007automatic} & 0.474 & 0.164 & 0.025\\
\small RigNet \cite{xu2020rignet} & 0.194 & 0.084 & 0.009 \\
\small  Ours & \textbf{0.012} & \textbf{0.007}  & \textbf{0.004} \\
\bottomrule
\end{tabular}
\label{tab:rig_quantitative}
\end{table}

\begin{figure*}[h]
    \centering
    \newcommand{\pll}{-1}
    \begin{overpic}[width=\linewidth]{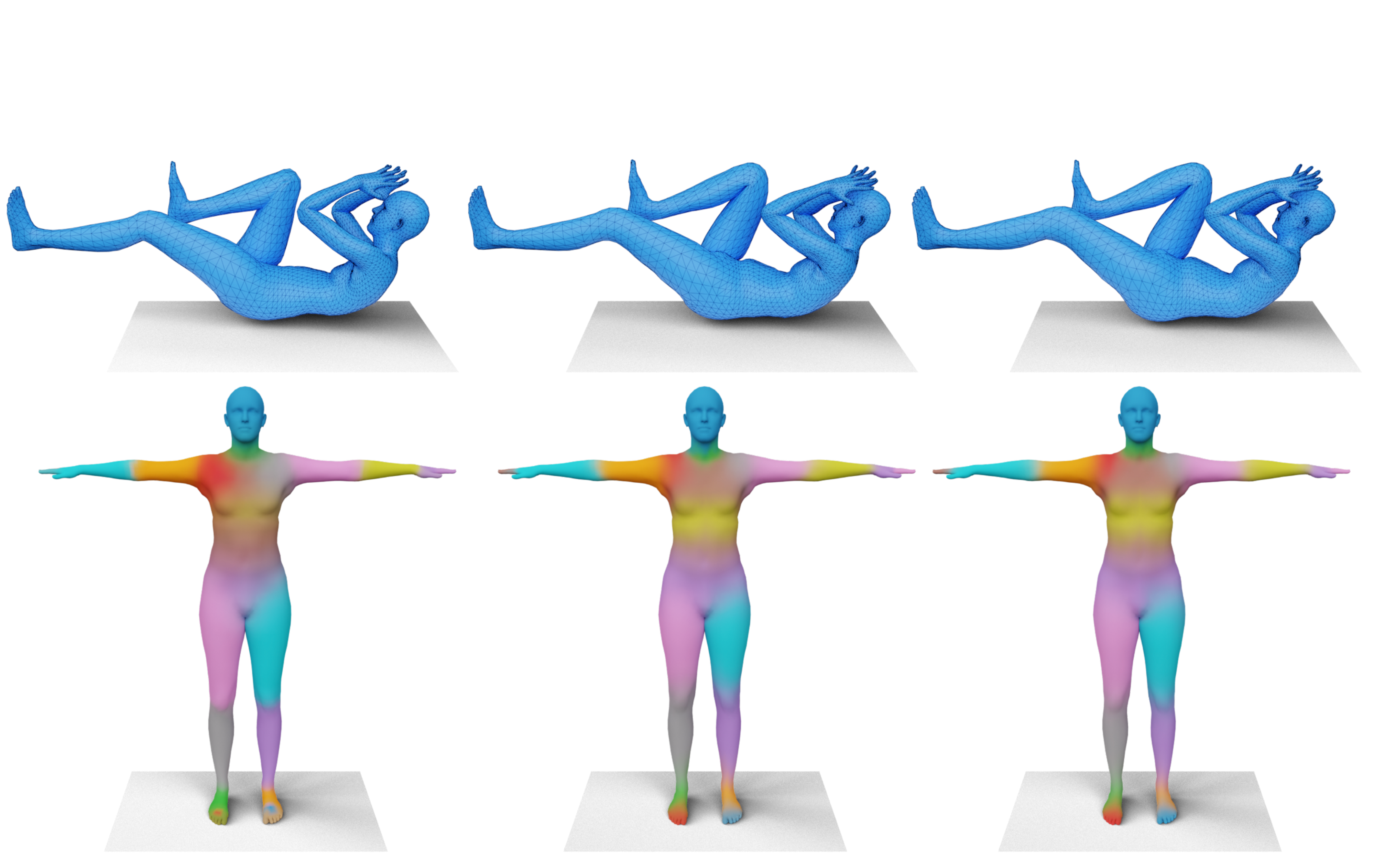}
    \put(5,  \pll){\textcolor{black}{Blender (based on Pinocchio~\shortcite{baran2007automatic})}}
    \put(49,  \pll){\textcolor{black}{Ours}}
    \put(77,  \pll){\textcolor{black}{Ground-truth}}
    \end{overpic}
    \caption{Comparison with the automatic skinning and deformation method in Blender (Based on Pinocchio~\shortcite{baran2007automatic}).}
    \label{fig:skincmp}
\end{figure*}

\paragraph{Skinning and Deformation}
We compared our skinning and deformation results to the output of Blender software which uses an updated version of the skinning algorithm from Pinocchio~\cite{baran2007automatic}. To measure the skinning error we use a simple L1 metric between the estimated skinning matrix and the ground-truth one. In order to perform this comparison we had to ensure that the number of joints is similar in all of the outputs, thus, we provided Blender with the ground-truth skeleton. In this way, we received skinning matrix with similar dimensions to the ground truth. 
For the deformation evaluation, we have chosen a fixed test sequence for each comparison and calculated the average error of vertex displacements and max error. The results are reported in Table~\ref{tab:skin_quantitative} and the output skinning matrix weights of each methods is visualized in Figure~\ref{fig:skincmp}. It can be seen that our method out performs the Blender software quantitatively and qualitative.

\begin{table}
\caption{Quantitative comparison between our skinning results to Pinocchio \cite{baran2007automatic}.}
\begin{tabular}{l c c c}
\toprule
& \small Skinning Weight(L1) &  \small Avg Dist. &  \small Max Dist. \\
\midrule
\small  Pinocchio \shortcite{baran2007automatic} & 23.1 & 0.27 & 20.3\\
\small  Ours & \textbf{2.56} & \textbf{0.011} & \textbf{1.68} \\
\bottomrule
\end{tabular}
\label{tab:skin_quantitative}
\end{table}

\subsection{Ablation study}
\paragraph{Residual Branch}
In this section, we perform an ablation study to evaluate the importance of each of the components in our system. In particular, we retrain our framework (a) without neural blend shapes (only envelope), (b) without connectivity augmentation, and (c) without garment augmentation. We calculate the L2 distance between the ground-truth displaced vertices and the predicted vertex displacements on the test dataset. The results are shown in Table~\ref{tab:ablation}. We can see that each of these components is important to our system, and removing any one of them will result in a drop in accuracy. Moreover, the most critical component which has the most impact on quantitative deformation accuracy is the mesh augmentation component. Although only using envelope (without using neural blend shapes) has the least numeric influence of the three, it is extremely important for a high visual quality deformation as we have showed throughout the paper.

\begin{table}
\caption{Ablation Study}
\begin{tabular}{c c c c}
\toprule
\small Envelope &  \small  No Mesh Aug &  \small No Garm Aug & All  \small  \\
\midrule
 0.024 & 0.66 & 0.23 & \textbf{0.011}\\
\bottomrule
\end{tabular}
\label{tab:ablation}
\end{table}

\paragraph{Connectivity Augmentation}
The network trained without connectivity augmentations produces undesirable deformation artifacts in the joints, and struggles to generate meaningful corrective blend shape displacements, which can be seen in the supplementary video and in Figure~\ref{fig:connectivity_ablation}. This shows that our connectivity augmentations are key to the generalization capabilities of the neural blend shapes on different mesh connectivity.
\begin{figure}
    \centering
\includegraphics[width=\linewidth]{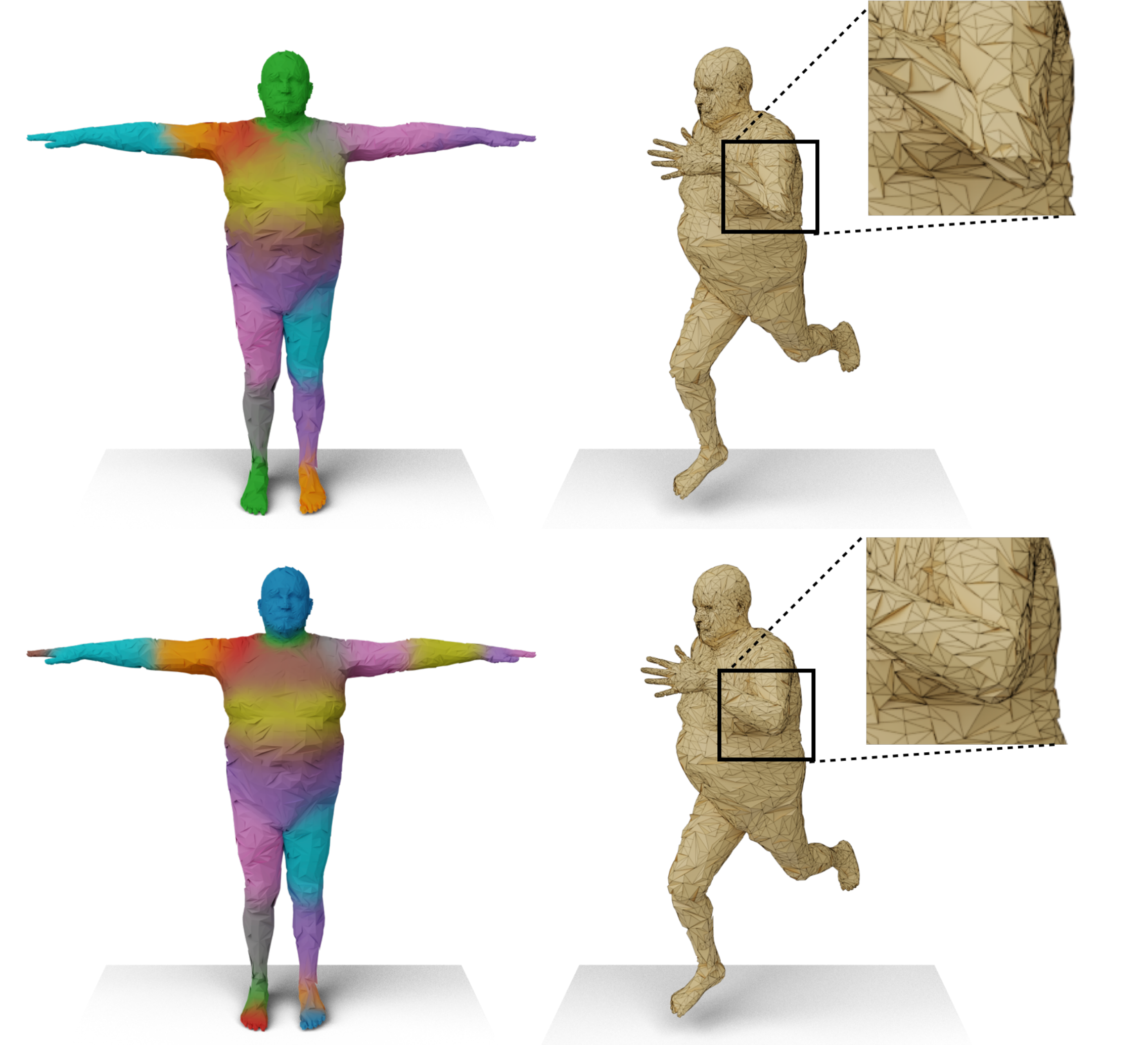}
    \caption{Connectivity augmentation ablation. 
    Top: network trained without connectivity augmentation, bottom: our network trained with connectivity augmentations. On the left are the network predicted skinning weights and on the right the deformed character. Observe that training with connectivity augmentations are key to generalization of the neural blend shapes on different mesh connectivity.}
    \label{fig:connectivity_ablation}
\end{figure}

\paragraph{Garment Augmentation}
The network trained without garment augmentations struggles to generalize to unseen characters.
This operation augments the input characters with piecewise smooth displacements extracted from the dataset of clothed characters and enables the network to not only generalize to clothed characters but also to achieve better results for naked, unseen characters. We believe this is because the garment augmentation creates a variety of geometric variability in the training data, especially with regard to the location of the skin w.r.t. the character bone.
In particular, we observe that the garment augmentation is critical factor in enabling our system to generalize the character created by a 3D artist, which can be seen in Figure~\ref{fig:garment_ablation} and in the supplementary video.
\begin{figure}
    \centering
\includegraphics[width=\linewidth]{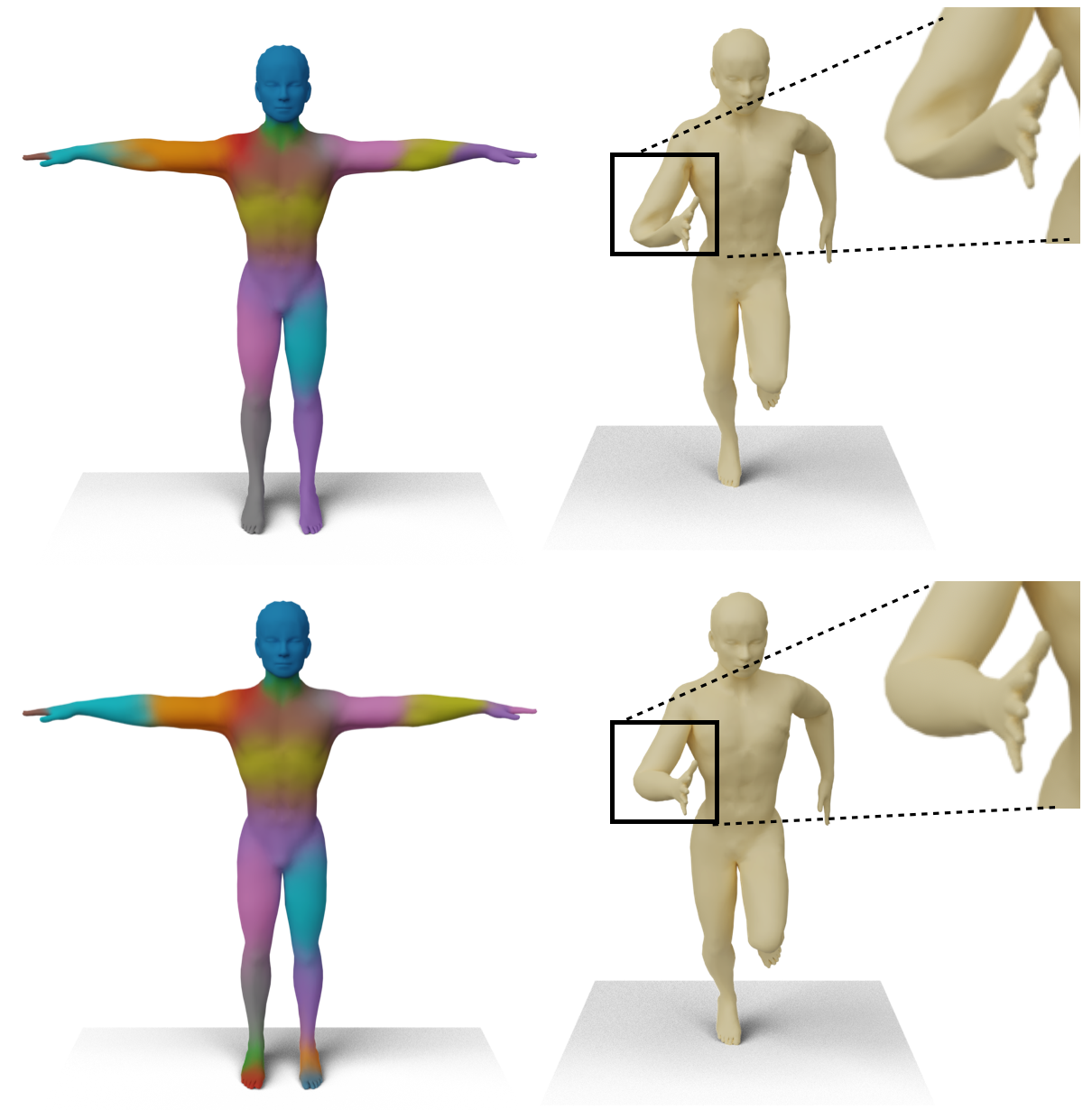}
    \caption{Garment augmentation ablation. Top: network trained without garment augmentation, bottom: our network trained with garment augmentations. On the left are the network predicted skinning weights and on the right the deformed character. Training with garment augmentations are key to our systems ability to generalize to the character created by a 3D artist.}
    \label{fig:garment_ablation}
\end{figure}

\paragraph{Envelope}
We trained our network only on the envelope deformation (\emph{i.e.,} without neural blend shapes). While our envelope obtains better deformations than LBS, it cannot produce pose-dependent corrective displacements in the joint regions. We can see that the quality of the deformation is improved when using the blend shapes residual corrective displacement. A result of this is shown in Figure~\ref{fig:bs_comparison}, where the volume of the mesh around the elbow are perfectly preserved even when the corresponding joints are transformed dramatically. Please refer to the supplementary video for more results.

\paragraph{Indirect Supervision}
Our network is trained with \textit{indirect supervision}, namely, ground-truth skinning and rigging samples are not provided during training, and the network is supervised only by the ground-truth deformation. This relaxes the assumption that the data samples should have a specific underlying deformation model. 
However, our training framework can easily employ direct supervision when ground-truth samples of skinning and rigging are available.
In such a case, we can apply $\ell_2$-loss to the difference between the generated skeleton and skinning weight to the ground-truth via
\begin{equation}
	\Loss_{\text{supervised}} = \Loss_{\text{s}} + \Loss_{\text{w}} = \| \tilde{O} - O\|^2 + \| \tilde{W} - W\|^2,
	\label{eq:supervised}
\end{equation}
where $\Loss_{\text{s}}$ and $ \Loss_{\text{w}}$ are the skeleton and skinning loss terms. 

As demonstrated throughout the paper, indirect supervision approach is capable to learn high deformation quality. However, as shown in Table~\ref{tab:direct}, direct supervision configuration converges faster comparing to the indirect supervision, although the deformation quality of both training configurations is comparable.

\begin{table}
\caption{Indirect supervision - Ablation study}
\begin{tabular}{l c c}
\toprule
\small  & Indirect Sup. &  Direct Sup. \\
\midrule
  $\Loss_{\text{w}}$ converges to & 0.14 & 0.10 \\
  $\Loss_{\text{s}}$ converges to & 0.012 & 0.009 \\
  \# of iters to reach $\Loss_{\text{v}} = 0.024$ & 80,000 & 45,000 \\
\bottomrule
\end{tabular}
\label{tab:direct}
\end{table}
\section{Discussion and conclusion}

We presented an approach to train a neural network to rig and skin an input character mesh  with a specific, prescribed skeleton structure and automatically generate neural blend shapes to enhance the articulated deformation quality in a pose-dependent manner. The fact that our framework incorporates the desired skeleton structure makes it practical for animation with existing motion data, such as available mocap libraries or legacy animation data of previously designed characters; the overall process is compatible with typical workflows in animation software. At the same time, the learned blend shapes ensure high quality of the output deformation, avoiding the usual LBS pitfalls, and are in general a powerful means to learn various deformation effects and apply them to unseen meshes with arbitrary connectivity. Unlike many existing example-based approaches, our system only requires a single mesh as input to compute the skeleton rig and the neural blend shapes, which makes it applicable to a large range of scenarios.

As is typical in the deep learning paradigm, the output of our method is bounded by the quality of the deformations that exist in the training dataset. 
Most of the results shown in this paper are trained using examples generated by SMPL-like models~\cite{loper2015smpl,bhatnagar2019mgn}. The resulting rigs give rise to high quality mesh deformation in most of our experiments, but can generate artifacts in certain cases where the selected model was not trained sufficiently. 
Fortunately, our indirectly supervised learning does not assume that the training data has a specific underlying model and can be easily extended to leverage other types of examples to learn versatile deformations. A direct enhancement to the results shown here is to train the system based on production rigs. We are also interested in exploring the possibility of learning secondary dynamics from simulation in the future.

Our system is trained to generate rigs with a prescribed skeleton structure. Changing this skeleton structure, \emph{e.g.,} adding extra joints or altering the connectivity, requires retraining the entire network. In future work, it would be interesting to explore skeleton-aware models, such as those proposed in~\cite{aberman2020skeleton}, to embed the skeleton structure in the rigging network to support arbitrary rigs using the same model, as well as relaxing some of the assumption we make on the input character (manifold mesh, T-pose shape, etc.).
Automatically adapting and combining multiple template skeleton structures to multi-component characters~\cite{miller2010frankenrigs,bharaj_automatically_2012} would be another interesting problem to explore.

\begin{acks}
We thank Sigal Raab and Andreas Aristidou for their valuable inputs that helped to improve this work. We also thank Cong Li and Yulong Zhang for their kind help with the experiments and ablation study. This work was supported in part by National Key R\&D Program of China (2019YFF0302902, 2020AAA0105200), Beijing Academy of Artificial Intelligence (BAAI), and Ant Group.
\end{acks}

\bibliographystyle{ACM-Reference-Format}
\bibliography{bibs}

\appendix

\section{Network Architectures}

\label{appendix}

In this section we describe the details for the network architectures. 

Table~\ref{tab:arch} describes the architecture for our envelope deformation branch and residual deformation branch, where \texttt{MConv, SConv, FC, LReLU, Pool} and  \texttt{Softmax} denote mesh convolution ~\cite{Hanocka2019MeshCNN}, skeleton-aware convolution ~\cite{aberman2020skeleton}, fully connected layer, leaky ReLU, max-pool $1/5$ channels and softmax activation.

\small{
\begin{table}
\begin{center}
    \begin{tabular}{ l l  l l }
    \toprule
    Name &  Layers  & i/o_c & LR\\
    \toprule \small{Skinning}
     & \texttt{MConv + LReLU + Pool}  & $3/64$ & $2 \times 10^{-4}$\\
     \small{Block}
     & \texttt{MConv + LReLU + Pool}  &  $64/128$ \\
     & \texttt{MConv + LReLU + Pool}  &  $128/256$\\
     & \texttt{MConv + Softmax} &  $256/24$\\

    \midrule \small{Geometry}
     & \texttt{MConv + LReLU}  & $3/64$ & $1 \times 10^{-5}$\\
     \small{Block}
     & \texttt{MConv + LReLU}  &  $64/128$ &  (stage 1) \\
     & \texttt{MConv + LReLU}  &  $128/256$ & $1 \times 10^{-4}$ \\
     & \texttt{MConv + LReLU} &  $256/512$ & (stage 2)\\
     
    \midrule \small{Offset}
     & \texttt{SConv + LReLU} & $512/256$ & $1 \times 10^{-4}$ \\
     \small{Block}
     & \texttt{SConv + LReLU} & $256/128$ \\
     & \texttt{SConv + LReLU} & $128/64$ \\
     & \texttt{SConv} & $64/3$\\
     
     \midrule \small{Blend Shape}
      & \texttt{MConv + LReLU} & $536/256$ & $1 \times 10^{-4}$ \\
      \small{Block}  
      & \texttt{MConv} & $256/27$ \\
      
     \midrule \small{Blend Shape}
      & \texttt{FC + LReLU} & $9/18$ & $1 \times 10^{-3}$ \\
      \small{Coefficient}
      & \texttt{FC + LReLU} & $18/32$ \\
      & \texttt{FC} & $32/9$ \\
    \bottomrule
    \end{tabular}
\end{center}
\caption{Network Architectures}
\label{tab:arch}
\end{table}}

\end{document}